# Electronic Trap Detection with Carrier-Resolved Photo-Hall Effect

Oki Gunawan[1,‡,*], Chaeyoun Kim[2,‡], Bonfilio Nainggolan[1,3], Minyeul Lee[2], Jonghwa Shin[2], Dong Suk Kim[4,5], Yimhyun Jo[4], Minjin Kim[4], Julie Euvrard[6], Douglas Bishop[1], Frank Libsch[1], Teodor Todorov[1], Yunna Kim[2], Byungha Shin[2,*]

1. IBM T. J Watson Research Center, Yorktown Heights, NY 10598 USA
2. Dept. of Materials Science and Engineering, Korea Advanced Institute of Science and Technology, Daejeon 34141, Republic of Korea
3. Dept. of Physics, Arizona State University, Tempe, AZ 85287, USA
4. Ulsan Advanced Energy Technology R&D Center, Korea Institute of Energy Research, Ulsan 44776, Republic of Korea
5. Graduate School of Carbon Neutrality, Ulsan National Institute of Science and Technology, Ulsan 44919, Republic of Korea
6. Dept. of Physics and Centre for Processable Electronics, Imperial College London, London SW7 2AZ UK

*Corresponding authors: ogunawa@us.ibm.com; byungha@kaist.ac.kr. ‡These authors contributed equally to this work.

**Abstract:**

Electronic trap states are a critical yet unavoidable aspect of semiconductor devices, impacting performance of various electronic devices such as transistors, memory devices, solar cells, and LEDs. The density, energy level, and position of these trap states often enable or constrain device functionality, making their measurement crucial in materials science and device fabrication. Most methods for measuring trap states involve fabricating a junction [1,2], which can inadvertently introduce or alter traps, highlighting the need for alternative, less-invasive techniques. Here, we present a unique photo-Hall-based method to detect and characterize trap density and energy level while concurrently extracting key carrier properties, including mobility, photocarrier density, recombination lifetime, and diffusion length. This technique relies on analyzing the photo-Hall data in terms of "photo-Hall conductivity" vs. electrical conductivity under varying light intensities and temperatures. We show that the photo-Hall effect, in the presence of traps, follows an *astonishingly simple* relationship — a *hyperbola equation* — that reveals detailed insights into charge transport and trap occupation. We have successfully applied this technique to *P* and *N*-type silicon as a benchmark and to high-performance halide perovskite photovoltaic films. This technique substantially expands the capability of Hall effect-based measurements by integrating the effects of the four most common excitations in nature — electric field, magnetic field, photon, and phonon in solids — into a single equation and enabling unparalleled extraction of charge carrier and trap properties in semiconductors.

The performance of semiconductor devices across a wide range of applications can be impacted by the presence of trap states. Traps in semiconductors originate from defects or impurities that capture and release free charge carriers. The effect of the traps can modify or dominate the electrical transport and charge carrier recombination process in many devices, often reducing conductivity or decreasing charge carrier lifetime. Traps can also give rise to hysteresis, noise in electrical screening or electrical leakage and unwanted power or state dissipation. The sources of traps include impurities and crystalline defects [3] such as point defects, extended defects (dislocations, grain boundaries and stacking fault), surface states and interface states. Understanding and controlling these traps is crucial for optimizing the performance of semiconductor devices. The origin of, and passivation of trap states has been extensively studied in silicon-based materials but continues to be an area of focus in optoelectronic devices such as solar cells and LEDs, as well as in small channel devices where interface states dominate transport. In other material systems, trap states are less explored but equally important. In particular, lead-halide perovskite systems have shown great promise for optoelectronic applications such as solar cells, recently surpassing the record power conversion efficiency of 26.7% in single-junction configurations [4-7]. The material system has shown inherently low trap density, enabling their high performance, however, managing and further reducing residual traps is increasingly critical for further performance improvements [8,9].

Techniques to detect traps in semiconductors include deep level transient spectroscopy (DLTS) [10,11], drive level capacitance profiling [1,12], space-charge-limited current [13,14], thermal admittance spectroscopy [2,15], transient photoluminescence [16], time-resolved microwave conductivity [17], photo-Hall spectroscopy [18], transient [19], and constant light photo-Hall effect [20]. Each technique offers unique advantages tailored to different aspects of trap properties measurement and suffers from some limitations. See Supplementary Information (SI) section E for a summary. In this study, we present a new technique based on recently-developed carrier-resolved photo-Hall (CRPH) effect [21] that addresses the major drawbacks of existing approaches including the need for a *p-n* or Schottky junction, the impact of interfaces associated with such device, and the transient nature of most measurements. This new carrier *and trap* resolved photo-Hall (CTRPH) technique also provides access to a rich set of charge carrier parameters of up to 17 charge carrier parameters that can be mapped against varying light intensities, including the electron and hole mobilities, photocarrier and trapped carrier density, recombination lifetimes, and diffusion lengths; and four parameters associated with the dominant trap state (density, energy level, and scattering cross sections). See SI section C for a complete list and descriptions.

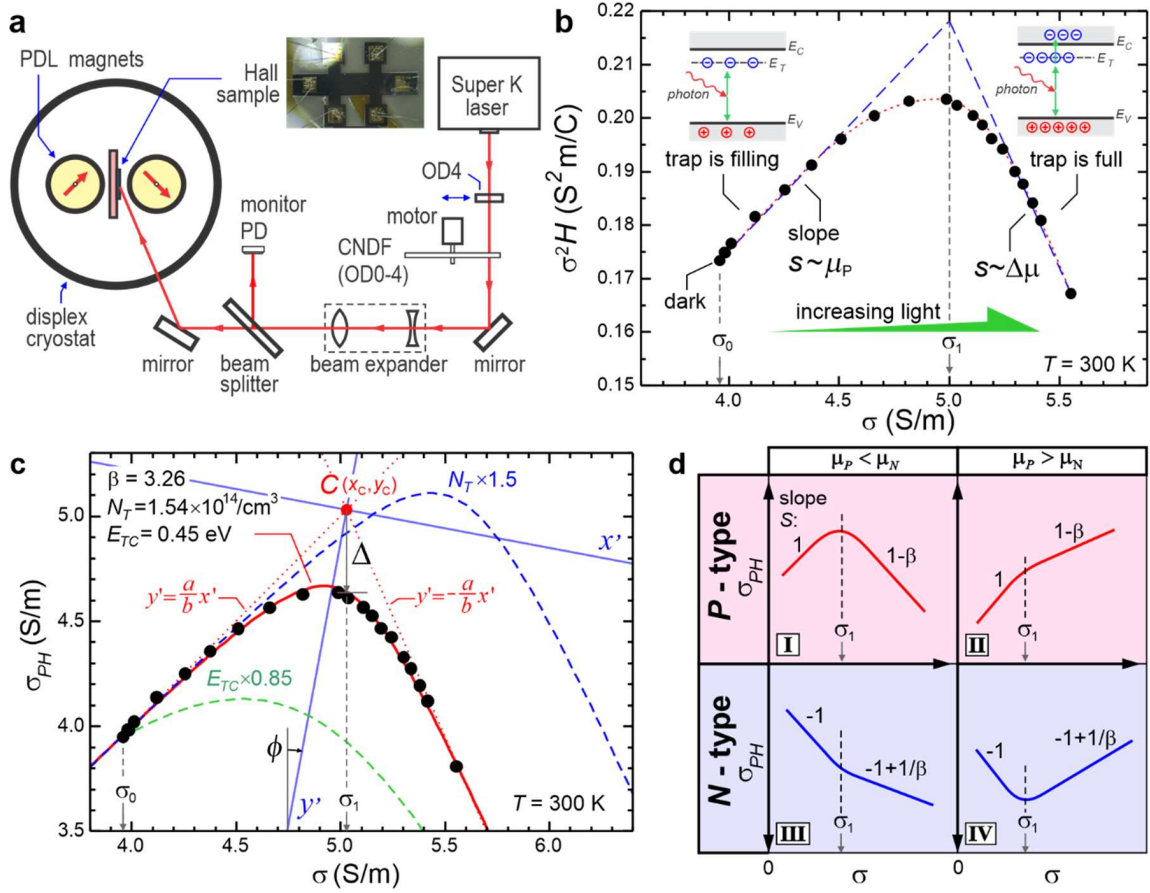

**Figure 1. Electronic trap detection with CTRPH technique.** (**a**) The experimental setup based on rotating PDL magnet Hall system with cryostat. PD is photodetector, OD is optical density filter and CNDF is continuous neutral density filter. **Inset:** the *p*-type SOI Hall sample (**b**) The photo-Hall experiment plot of $\sigma^2 H$ vs. $\sigma$ in a *p*-type SOI sample with laser light of wavelength $\lambda$=615 nm, maximum intensity $I_L$=157 mW/cm$^2$, and temperature $T$=300 K. The dashed curve is only a guide to the eye. (**c**) Hyperbolic fit (red curve) of the data in Fig. 1b. The *x'-y'* is the principal axis of the hyperbola centered at point *C*. The value of fit parameters $\beta$, $N_T$ and $E_{TC}$ are indicated. The impact of varying $N_T$ and $E_T$ are simulated in the blue and green dashed-curve respectively. (**d**) Four possible scenarios (marked as quadrant I-IV) of photo Hall behavior with trapping effect depending on carrier type (*P* or *N*) and the relative value of the hole and electron mobility. The dashed lines separate the first and second regime of the curves.

The experimental setup is shown in Fig. 1a. The Hall sample is mounted inside a cryostat with temperature control ranging from 20 K to 340 K. The sample is placed between two rotating parallel dipole line (PDL) magnets that generate a strong perpendicular oscillating magnetic field [22,23] with peak amplitude of $B\sim0.5$ T. We use a supercontinuum white laser that passes through a monochromator and a set of optics to control light intensity and beam size. To demonstrate our trap detection technique, we use a *p*-type silicon-on-insulator (SOI) sample on glass with thickness $d$=5.0 μm. The device is a six terminal Hall bar (see Fig. 1a inset). The detailed experimental setup and sample fabrication are described in Methods.

The measurement is similar to previously demonstrated CRPH technique [21] – in brief, we measure the electrical conductivity $\sigma$ and Hall coefficient $H$ with increasing light intensity and plot $\sigma^2 H$ vs. $\sigma$ [21] as shown in Fig. 1b. In our previous work [21], we demonstrated that the minority and majority carrier mobilities can be extracted from the slope $s = d(\sigma^2 H)/d\sigma$ which yield the difference between the hole and electron mobility, i.e. $s = \Delta\mu_H = r(\mu_P - \mu_N)$, where $r$ is the Hall scattering factor and is assumed to be ~1 [24]. This model applies when the trapping effect is negligible (as in high quality materials or high light intensity), allowing us to assume that the electron and hole photocarrier density are equal ($\Delta n = \Delta p$). If we extend this model by considering trapping effect, where all the minority carriers are trapped, we show that the slope will yield the majority carrier mobility $\mu_0$ i.e. $s \sim \mu_0$ (see SI section A.1 for derivation).

The key insight of the trap detection using CTRPH technique can be seen in the $\sigma^2 H - \sigma$ trace where there is a bending that demarcates two regimes of the plot with positive and negative slope at low and high light intensity respectively. If we draw two asymptotic lines as shown in Fig. 1b we obtain slope of $s$ = 440 and -990 cm²/Vs respectively, which is consistent with the known mobility values of hole (the majority carrier) in silicon: $\mu_P$ ~ 500 cm²/Vs and mobility difference $\Delta\mu = \mu_P - \mu_N$ ~ −1000 cm²/Vs [3]. We can understand this behavior by considering trapping effect of the minority carrier as shown in Fig. 1b inset. At low light intensity, the photo-generated minority carriers initially populate the traps, and only free photo-generated majority carriers ($\Delta p$) contribute to the conductivity, thus $s \sim \mu_P$. Once all the traps are filled at certain light intensity, the photo-generated minority carriers ($\Delta n$) will populate the conduction band (CB) and start contributing to the conductivity. In the limit where the photo-generated carrier density $\Delta n$ and $\Delta p$ are much higher than the trap density $N_T$, we obtain $\Delta n \sim \Delta p$ and $s \sim \Delta\mu_H$ as observed in our earlier study where trapping effects are negligible [21,25].

We now proceed to a more quantitative analysis of the trap properties. To describe the photo Hall effect in the presence of traps, we model a single level trap in a $P$-type material with density $N_T$ and energy $E_T$ which is closer to CB thus acting as minority carrier trap (see SI section A.1). As the light intensity is increased, the electron quasi-Fermi level increases and the trapped ($n_T$) and free ($\Delta n$) electron density can be calculated. To simplify our analysis, we define a new quantity: $\sigma_{PH} = \sigma^2 H / r\mu_0$, which we refer to as "photo Hall conductivity" and shares the same dimension as $\sigma$. We can show that the $\sigma_{PH}$ vs. $\sigma$ plot in the presence of traps can be described by a *simple hyperbola equation* (see SI section A.2 for derivation):

$$\frac{\sigma_{PH}^{'2}}{a^2} - \frac{\sigma^{'2}}{b^2} = 1 \tag{1}$$

where ($\sigma'$, $\sigma_{PH}'$) are ($\sigma$, $\sigma_{PH}$) expressed in a second coordinate system x'-y' which is centered at point C and rotated by angle $\phi$ as shown in Fig. 1c. We have $\phi = \frac{1}{2}\tan^{-1}(2/\beta - 1)$, where $\beta = \mu_N / \mu_P$ is the electron-to-hole mobility ratio. To describe the hyperbola equation parameters, we define another two parameters: $s_N = e\mu_0(p_0 + N_T)$ and $s_E = e\mu_0(1+\beta)N_C \exp(-\varepsilon_{TC})$ where $\varepsilon_{TC} = (E_C - E_T)/k_B T$ with $E_C$ is the CB edge, $e$ is the electron's charge, $p_0$ is the majority carrier density in the dark, $N_T$ is the trap density, $N_C$ is the CB effective density of states, $k_B$ is the Boltzmann constant, and $T$ is the temperature. The parameter $s_N$ and $s_E$ have clear physical and geometrical meaning: $s_N$ depends on $N_T$ and determines the horizontal peak position ($\sigma_1$) of the hyperbola; $s_E$ depends on $E_T$ and determines the vertical peak position of the hyperbola or the vertical gap "$\Delta$" from the center point C (see Fig. 1c). They are related as: $s_E = \Delta^2 / e\beta^2 \mu_0 N_T$.

The hyperbola has a center point C with coordinate: $x_C = s_N - s_E$ and $y_C = s_N + (\beta - 1)s_E$. We also have $a$ and $b$ as the hyperbola semi major and minor axis respectively, where: $a = \sqrt{-K/\lambda_-}$ and $b = \sqrt{K/\lambda_+}$, with $K = e\mu_0 \beta^2 N_T s_E$ and $\lambda_\pm = \left(\beta - 2 \pm \sqrt{2\beta^2 - 4\beta + 4}\right)/2$ as the eigen values of the hyperbola matrix (see SI section A.2). We note a special case when $\beta = 2$, we have $\phi = 0$, or in other words, the hyperbola x'-y' axes are already aligned with the original $\sigma - \sigma_{PH}$ axes, thus no rotation is needed. From Eq. 1, we can also express $\sigma_{PH}$ explicitly as a function of $\sigma$ (see SI section A.1):

$$\sigma_{PH}(\sigma) = \left[\beta(s_N + s_E) + (2-\beta)\sigma - \beta\sqrt{(s_N + s_E + \sigma)^2 - 4(s_E\sigma_0 + s_N\sigma)}\right]/2, \tag{2}$$

with $\sigma_0$ is the conductivity in the dark. Eq. 2 can be used for curve-fitting the experimental data to extract the parameters $\beta$, $s_N$, $s_E$ and thus $N_T$ and $E_T$. This approach was applied in Fig. 1c, which also presents the resulting fitting parameters.

We note a few important features of the hyperbolic curve as described by Eq. 1 and 2. The slope $d\sigma_{PH}/d\sigma$ approaches 1 in the dark at low temperature (or $E_{TC} \gg k_B T$) and 1-$\beta$ at maximum light intensity. The angle $\phi$ and the asymptotes: $y' = \pm(b/a)x'$ are solely determined by mobility ratio $\beta$. The parameters $N_T$ and $E_T$ determine the horizontal and the vertical position of the inflection (or vertex) point of the hyperbola respectively. At larger $N_T$, the inflection point occurs at higher light intensity (or conductivity), as shown in the blue dashed curve in Fig. 1c where $N_T$ is increased by 1.5×. This shift is expected as more photo-generated carriers are required to fill the traps. Meanwhile, smaller trap energy $E_{TC}$ leads to a smoother hyperbolic curve as shown in the green dashed curve in Fig. 1c, where $E_{TC}$ is reduced to 0.85×. As the impact of the trap on the minority carrier is temperature-dependent, shallow traps ($E_{TC} \sim k_B T$) produce smoother hyperbolic curves, while deep traps ($E_{TC} \gg k_B T$) results in sharper hyperbolic curves. Fig. 1d. summarizes the expected shape of the $\sigma_{PH} - \sigma$ trace for four possible scenarios (marked as quadrant I to IV) depending on the material types (P or N) and the relative mobility values ($\mu_P < \mu_N$ or $\mu_P > \mu_N$). Our hyperbola model (Eq. 1 and 2) applies for all cases, albeit with minor modifications in formula for N-type systems (see SI section A.5). This diagram also reveals an important characteristics: significant bending behavior occurs when the minority carrier mobility is larger than that of majority carrier (Fig. 1d, quadrant I and IV). For example, the p-SOI data in Fig. 1c is consistent with quadrant I behavior.

We can use two approaches to extract the trap parameters $N_T$ and $E_{TC}$ from the photo Hall experiment. One technique would require to fit the experimental data ($\sigma_{PH} - \sigma$) using Eq. 1 or Eq. 2, and extract $\beta$, $s_N$ and $s_E$ as fitting parameters. This curve fitting is illustrated in Fig. 1c, that yields: $\beta = 3.26$, $N_T = 1.54 \times 10^{14}$ /cm$^3$ and $E_{TC} = 0.45$ eV. The $\beta$ value is close to 3 as expected for silicon [3]. This approach is however hindered by the need for effective mass that appears in the effective density of states $N_C$ (or $N_V$) in calculation involving $s_E$, which, while known for many materials, is rarely available for new materials.

In the second technique, which is an alternative to curve-fitting, we can perform a geometric analysis of the $\sigma_{PH} - \sigma$ trace to extract $\beta$, $N_T$ and $E_{TC}$. $\beta$ is extracted from the slope of the high light intensity regime with the asymptotic limit of 1-$\beta$, and calculated using: $\beta = 1 - (d\sigma_{PH}/d\sigma)_\infty$. Note that $\beta$ is however not required to calculate $N_T$ and $E_{TC}$. When $E_{TC} \gg k_B T$, which can be achieved at low temperature, the hyperbola becomes sufficiently sharp and the trap density can be estimated with a very simple equation (see SI section A.3):

$$\tilde{N}_T = \frac{\sigma_1 - \sigma_0}{e \mu_0} \tag{3}$$

where $\tilde{N}_T$ is the estimated trap density calculated from the difference between the inflection (or vertex) point at $\sigma_1$ and the dark conductivity $\sigma_0$ as shown in Fig. 1c. The difference in conductivity between the point $\sigma_1$ and $\sigma_0$ is primarily due to free majority carriers ($\Delta p$). As all available trap states are filled by the photo-generated electrons near the inflection point of the hyperbola, the $\Delta p$ at this point is close to the trap density $N_T$. A more exact expression of Eq. (3) that contains temperature correction is given in SI section A.3.

Next we can determine the trap energy $E_{TC}$ by measuring the gap "$\Delta$" in Fig. 1c and use the relationship (see SI section A.4 for derivation):

$$E_{TC} = k_B T \ln\left(e^2 \mu_0^2 \beta^2 (\beta+1) N_C N_T / \Delta^2\right). \tag{4}$$

In this example, we have $\Delta=0.462$ S/m and thus $E_{TC} = 0.46$ eV, consistent with direct curve-fitting technique in Fig. 1c.

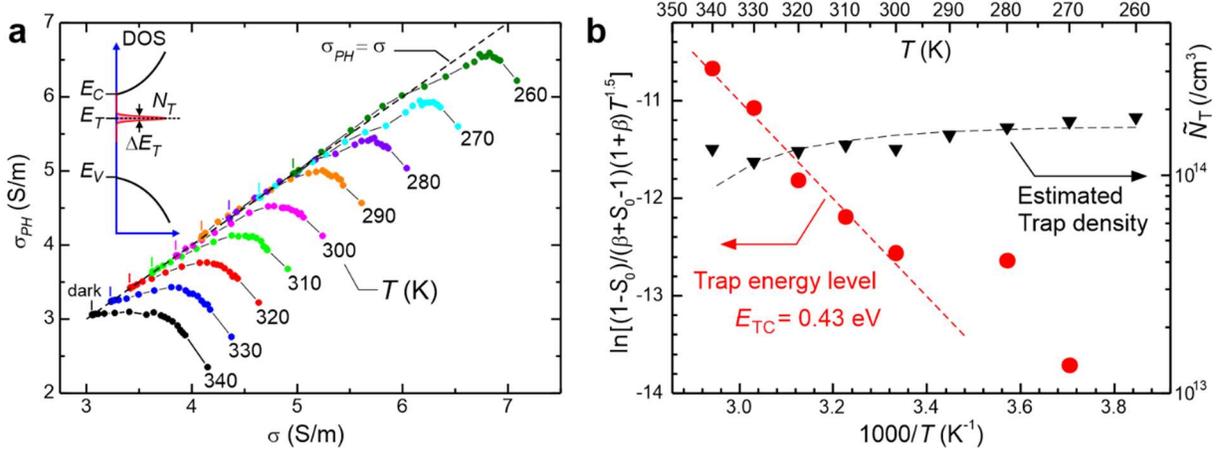

Fig. 2 **Trap analysis using temperature-dependent CTRPH measurement in the *p*-SOI sample. (a)** Photo-Hall data from $T=260$ to $340$ K with laser light of $\lambda=550$ nm and maximum intensity $I_L=165$ mW/cm$^2$. The dark data points are indicated by short vertical dashes. The dashed line is the $\sigma_{PH} = \sigma$ asymptotic line. **Inset:** energy band diagram of the trap model. **(b)** Arrhenius plot of Eq. 4 to extract the trap energy level ($E_{TC}$) and the estimated trap density ($\tilde{N}_T$) extracted using Eq. 3. Black dashed-curve: curve-fit of temperature-dependent effect of $\tilde{N}_T$ (see text).

In addition, we can also perform variable temperature photo-Hall measurements which can provide more accurate access to $E_{TC}$ without the need for prior knowledge of effective mass or the center point $C$. We demonstrate that the slope $S_0 = [d\sigma_{PH}/d\sigma]_0$ at the dark point is related to $E_{TC}$ according to (see SI section A.4):

$$\ln\left(\frac{1-S_0}{(\beta+S_0-1)(1+\beta)T^{1.5}}\right) = c_0 - \frac{E_{TC}}{k_B T} \tag{5}$$

where $c_0$ is a constant. Here, we rely on the fact that the slope $S_0$ drops significantly towards higher temperature as shown in Fig. 2a. We can perform an Arrhenius analysis by plotting the left-hand-side quantity vs. $1/T$ and extract $E_{TC}$ from the slope. We apply this variable temperature approach to the photo Hall measurements of the *p*-SOI sample at $T=260$ K to $340$ K as shown in Fig. 2. We observe that all curves show hyperbolic behavior and become sharper at lower temperature or $E_{TC} \gg k_B T$. The first segment of the curve approaches asymptotic limit: $\sigma_{PH} = \sigma$ towards low temperature as expected from the theory. We obtain trap density $N_T = 1.66 \times 10^{14}$ /cm$^3$ from the hyperbola peaks and using Eq. 3 at low temperature range ($T \sim 260$ K). This value is close to $N_T = 1.54 \times 10^{14}$ /cm$^3$ determined from curve-fitting approach in Fig. 1c. We also observe in Fig. 2b, the $\tilde{N}_T$ values drop slightly at higher temperature which is consistent with the theoretical prediction (indicated as black-dashed curve) described in SI section A.3.

Using Eq. 5, we extract $E_{TC}$ from the slope of the data in Fig. 2b. A linear behavior with respect to $1/T$ is observed at high temperature range (300-340 K), the slope yields $E_{TC} = 0.43$ eV. At low temperature ($T < 300$ K) the slope $S_0$ is approaching 1 and the left-hand-side quantity in Eq. 4 becomes less accurate and deviates from the Arrhenius-behavior trend (Fig. 2b) and thus the data in this temperature range can be excluded. This $E_{TC}$ value is very close to the value obtained from previous technique using curve-fitting or "Δ" calculation, i.e. (0.45 eV). Some discrepancy from the first method can be attributed due to inaccuracy of point $C$ determination (that relies on intersection of two asymptotes) and the value of the effective mass of the SOI sample (which could change under strain). Nevertheless, close and consistent results of $N_T$ and $E_{TC}$ demonstrate the reliability of both approaches. We note that similar trap level of 0.48 eV has been previously observed in SOI samples using DLTS study, and the trap is suspected due to small voids from poorly bonded SOI [26].

To complete the analysis of the SOI sample, we extend our CTRPH analysis in similar fashion with our previous work [21], but now incorporating the presence of traps (see SI section B). This approach enables us to fully resolve the concentration of electrons in the trap ($n_T$), electron photocarrier density ($\Delta n$) and hole photocarrier density ($\Delta p$) using: $n_T = [(\beta-1)\sigma + \sigma_{PH} - \beta\sigma_0] / e\mu_P\beta$, $\Delta n = (\sigma - \sigma_{PH})/e\mu_P\beta(\beta+1)$ and $\Delta p = \Delta n + n_T$ (see SI section A.1). Consequently, we can now calculate up to 17×N charge carrier parameters, where N represents the number of light intensity settings. A complete list is provided in SI section C. The electron and hole charge carrier parameters include: photo carrier densities ($\Delta n$, $\Delta p$), occupied trap density ($n_T$), mobility ($\mu_N$, $\mu_P$), recombination lifetime ($\tau_N, \tau_P$), diffusion coefficients ($D_N$, $D_P$, $D_A$) and diffusion lengths ($L_{D,N}, L_{D,P}, L_{D,A}$), where subscript N, P, A mean electron, hole, and ambipolar quantity respectively. Given the carrier density information, we can also calculate quasi-Fermi levels ($QF_N$, $QF_P$), their splitting (QFLS) and its ideality factor ($\eta$) as suggested in Ref. [20]. Additionally, we can extract four trap parameters, including trap density $N_T$, trap energy $E_T$ and recombination scattering cross sections ($\sigma_P, \sigma_N$). The complete extraction of these parameters, along with relevant plots for the p-SOI sample, is presented in SI section D.2. We have also demonstrated CTRPH analysis in an N-type silicon sample, which exhibits quadrant III behavior in Fig. 1d, as detailed in SI section D.3.

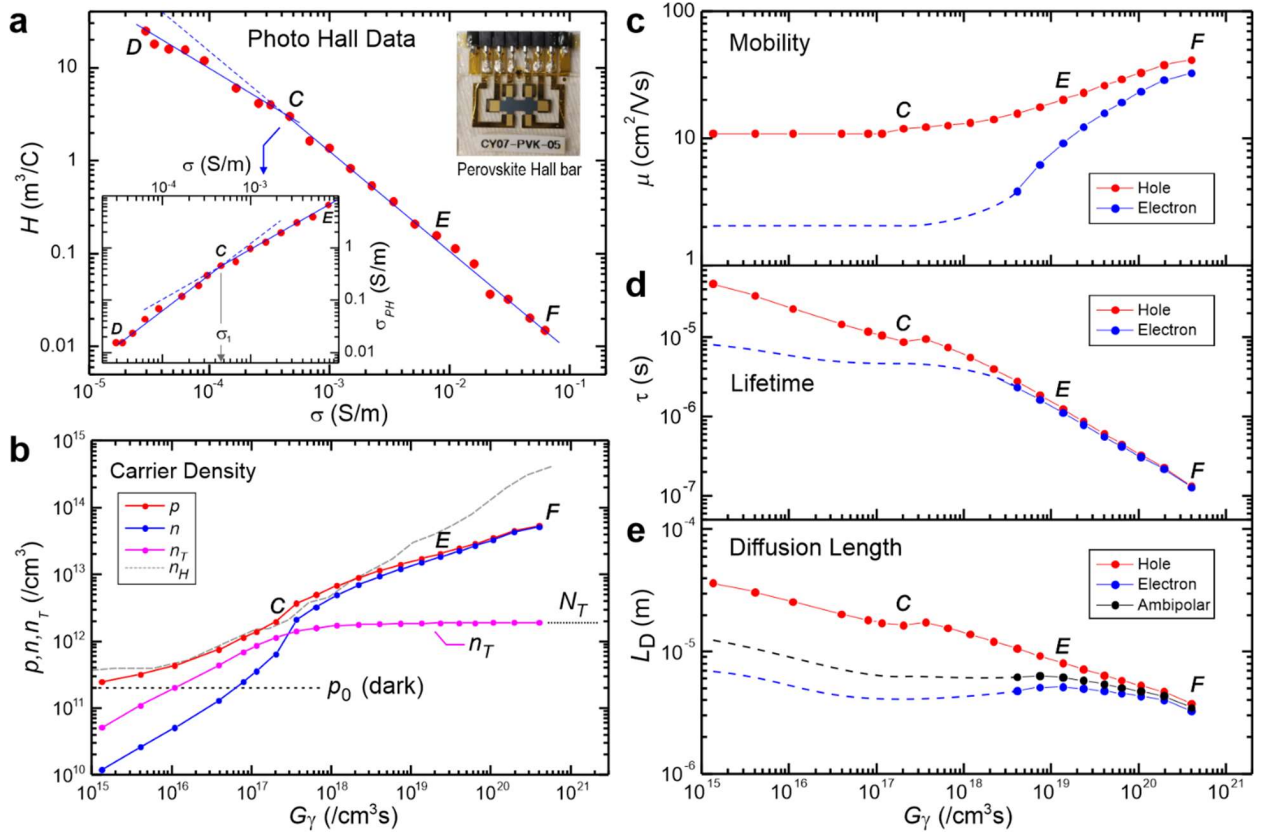

Fig. 3 **CTRPH trap analysis of a high-performance perovskite film. (a)** The photo Hall data with laser light λ= 615 nm and maximum intensity $I_L$ =16 mW/cm². Point *D*, *C*, *F* mark the dark, inflection point, and full light intensity respectively. **Lower Inset:** $\sigma_{PH}$ vs. $\sigma$ data that shows bending behavior due to trap. **Upper inset:** the perovskite Hall bar sample. **(b)** Carrier density vs. absorbed photo density ($G_\gamma$). $n_H = 1/eH$ is the Hall density for comparison. **(c)** Mobility. **(d)** Recombination lifetime. **(e)** Diffusion length. Dashed curves in **c**, **d**, **e** are estimated minority carrier values as the minority mobility values between dark and inflection point *C* are more uncertain.

We further apply the CTRPH technique to a high-performance FAPbI$_3$ perovskite photovoltaic film, processed identically to the perovskite layers used in solar cell devices that recently achieved a record power conversion efficiency of 25.4% [27]. The sample is a six-terminal Hall bar on glass with thickness: *d* = 0.7 μm. The sample demonstrates a CTRPH trap analysis in a unique context where: (1) $\mu_P > \mu_N$ (Fig. 1d, quadrant II), (2) a high injection regime ($\Delta p \gg p_0$), and (3) variable mobility at high light intensity. To address this characteristics, the analysis is divided into two regimes: trapping regime (indicated by data point range *D-C-E* in Fig. 3a) and high light intensity where the mobility may vary (range *E-F*). In the trapping regime, we observe that the perovskite data exhibits the quadrant II (Fig. 1d) behavior where $\mu_P > \mu_N$. From the inflection point *C* in Fig. 3a inset, we determine the trap density: $N_T$ = 1.9×10$^{12}$ /cm³ using Eq. 3, which is ~30× smaller than $\Delta n$ or $\Delta p$ at maximum light intensity (see Fig. 3b). Notably, this $N_T$ value is significantly below the detection limit of junction-capacitance-based trap extraction methods ($N_{T,min} \sim 10^{15}$ /cm$^3$ at $d$=0.7 μm) [9,28], underscoring the advantage of this technique. The hyperbola curve profile is however too smooth to confidently determine the gap $\Delta$ and the trap energy level, but we estimate $E_T \sim$ 0.3-0.7 eV.

In the high light intensity regime, we obtain self-consistent solutions using an iterative technique for intensity-dependent mobility, as described in our previous work [21] (see also SI section D.4 for more detailed analysis). The extracted parameters are displayed as a function of absorbed photon density $G_\gamma$ in Fig. 3b-e. We observe that this perovskite sample exhibits excellent charge carrier parameters ($\mu \sim$ 40 cm$^2$/Vs, $\tau \sim$ 100 ns and $L_D \sim$ 3 μm) at maximum light intensity of ~0.16 sun, along with a very low trap density $N_T$. This is consistent with the recent record performance of the associated solar cell device [27].

We note another aspect of this work, our technique employs a simple trap model, assuming a single dominant trap with a narrow energy level, that governs the trapping effect and transport behavior in the material. In practice, multiple traps or more complex situations – such as distributed traps, tail states or Urbach tail – may exist, accompanied by a distribution of carrier mobilities. In principle, our CTRPH analysis can be extended to address these cases, though it would require more elaborate numerical computation.

In summary, we have demonstrated a comprehensive approach for extracting trap and charge carrier properties from semiconductors using the CTRPH technique, which relies on a detailed analysis of the $\sigma_{PH}$ vs. $\sigma$ curve, whose underlying behavior follows a newly-derived hyperbola equation. This work provides a unified framework for gaining deeper insights into charge carrier transport and trapping effects by using the most common excitations: electric field, magnetic field, photon (light), and phonon (via lattice temperature control) based on the Hall effect. Beyond traditional electric and magnetic field applications in Hall measurements, increasing photon intensity allows us to fill traps, determine their density, and evaluate the mobility ratio of the charge carriers. By varying temperature or phonon excitation, we probe the trap energy levels. This technique significantly enhances the capabilities of Hall effect measurements and holds potential for broad application, enabling single-experiment, comprehensive evaluation of charge carrier transport and trap properties in many emerging advanced electronic materials.

## Methods:

**Experimental Setup**

The setup is shown in Fig. 1a. The PDL Hall system consists of a dry cryogen-free displex cryostat system with a custom-built cold stage equipped with a 50W resistive heater, enabling temperature control from 20K to 340K using a Lakeshore 340 temperature controller. The fundamental operation of the PDL Hall system is detailed in Refs. [21-23]. The sample is mounted on the cold stage, positioned between two rotating PDL magnets made of NdFeB, each measuring 25.4mm in diameter and 25.4 mm in length. The magnetic field applied to the sample is $B=\pm0.5T$.

Light illumination is provided by either a red-green-blue (RGB) laser or supercontinuum white laser (Super K Fianium FIU-15 from NKT Photonics) equipped with monochromator system. Light intensity is controlled over eight orders of magnitude using a combination of a continuous neutral density filter (CNDF) and optical density 4 (OD4) filter. The CNDF is driven by a stepper motor and the OD4 filter is driven by a two-position motorized flipper. A series of biconvex and biconcave lenses expand the beam to obtain ~10 mm coverage diameter to ensure uniform illumination of the entire Hall sample. The light beam is then split by a beam splitter, with one part is directed downward to a "monitor photodetector" for real-time intensity monitoring and the other is directed at the sample.

The electronic instrumentation system consists of a Keithley 2450 source meter unit for providing current or voltage to the sample, a Keithley 2182A nanovoltmeter for measuring longitudinal and transverse voltage across the Hall sample, and a Keithley 6485 picoammeter to measure photocurrent or light intensity via the monitor photodetector or the reference photodetector cell. We have also built a custom PDL electronic control box that contains Raspberry Pi (RPi) Compute Module 4 microcomputer, a four-channel stepper motor controller (EZ4AXIS23WV from AllMotion), Pi-Plate data acquisition boards with analog and digital I/O mounted on top of the RPi and custom-designed Hall effect switch matrix unit. This high-input impedance switch matrix (capable of resistance measurements up to 1 TΩ) features six sample input channels, two source input channels, and two voltage read-out channels, allowing simultaneous measurements of longitudinal ($R_{xx}$) or transverse ($R_{xy}$) resistance.

The stepper motor controller controls the rotation of the PDL magnets, the CNDF and a linear stage for switching between the RGB laser and the Super K laser. The RPi microcomputer functions as a server, controlling operations of various modules inside the box. The box is controlled by a client Windows computer that runs MATLAB programs to execute various measurement operations.

**Fabrication of SOI Hall samples**

The sample is an intrinsic SOI wafer, consisting of a 5 μm single-crystalline intrinsic Si device layer, a 500 nm $SiO_2$ layer and a 525 μm Si handle layer. To transfer this single-crystalline Si film onto a $SiO_2$ substrate, the SOI wafer was bonded to a 500 μm thick borosilicate glass using a fusion bonding process at 673 K and a pressure of 1.5 kN. Subsequently, the unwanted handle layer was removed via chemical mechanical polishing process, and residual Si was etched using a $SF_6$ dry etch process. The remaining 500 nm $SiO_2$ layer was then chemically etched with HF solution, yielding the single-crystalline Si film on the $SiO_2$ substrate wafer. This $Si/SiO_2$ wafer was then diced into 10 mm×10 mm pieces, and a photolithography process was used to form a Hall bar photoresist pattern on the Si film. The Si Hall bar was then created using $SF_6$ dry etching, followed by photoresist removal. Finally, 200 nm thick Au electrodes were deposited on the Hall bar using an e-beam evaporator. The SOI Hall bar sample was mounted on a custom-made sample printed circuit board (PCB) package using "GE" varnish, and the Hall bar contacts were wire-bonded to the PCB pads. For variable temperature photo-Hall measurements, a small diode temperature sensor was attached to the glass substrate to accurately monitor the sample's temperature.

**Fabrication of perovskite Hall samples**

The perovskite Hall samples are based on formamidinium-rich lead iodide ($FAPbI_3$) films with methylammonium chloride (MACl) additives. This high-quality perovskite exhibits a solar cell device power conversion efficiency of 25.7% and has a bandgap of 1.55 eV as reported in Ref. [27,29]. Following these reports, $FAPbI_3$ black powder was synthesized by first dissolving $PbI_2$ and FAI in a 1:1 molar ratio in 2ME (0.8 M), and then filtered using a polyvinylidene fluoride filter with 0.45 μm pore size. The filtered solution was placed in a flask incubated in an oil bath at 120 °C for 1 h with slow stirring. The resulting black powder was filtered using a glass filter and dried at 60 °C for 24 hrs.

The perovskite precursor solution was prepared by dissolving 1550 mg $FAPbI_3$ and 61 mg MACl with 1mL DMF/DMSO (4:1). The perovskite solution was filtered with a polyvinylidene fluoride filter (0.2 μm) and then 70 μL of the filtered perovskite solution was spread onto the 2.5 cm×2.5 cm glass substrate at 8000 rpm rotation speed for 50 s. During spin-coating, 1 mL diethyl ether was dropped on the perovskite film at 10 seconds using a home-made pipette. The resulting film was annealed at 150 °C for 15 mins, followed by 100 °C for 30 mins on a hotplate, yielding high quality $FAPbI_3$ film with a thickness of 700 nm was obtained.

To perform photo-Hall measurement, a six-terminal perovskite Hall bar device was patterned by scraping away the excess film outside the Hall bar region. A 100 nm thick Au contact pattern was then deposited on the film. An eight-terminal header pin was attached to the glass substrate with epoxy, and its metal pins were connected to the Au contact pattern using silver epoxy. Since the perovskite thin film is highly sensitive to air and moisture, the Hall bar was encapsulated by attaching a secondary glass cover with epoxy. A small diode temperature sensor was also attached to the sample to monitor its temperature.

## Acknowledgements

O.G. acknowledges IBM Exploratory Research program. C.K. and B.S. acknowledge the financial support from the National Research Foundation of Korea (NRF) grants funded by the Korea government (MSIT, No. RS-2023-00208832 and No. NRF-2022M3H4A1A03074093). C.K and B.S acknowledge the financial support of BK21FOUR from R&E Initiative for K-Materials Global Innovation. J.E. acknowledges Royal Society Research Grants 2023 Round 2-RG/R2/232246. We thank Seung Ju Choi of Ulsan Advanced Energy Technology R&D Center, KIER and Sanghyuk Ryu of KAIST for assistance in preparing the perovskite samples and Michael Pereira for machine fabrication of the IBM PDL Hall system.

## Author contributions

O.G. (IBM) and B.S. (KAIST) served as principal investigators and conceived the project. O.G. led the project, constructed the experimental setup, programmed analysis codes, performed measurements, developed analysis and theory. B.S. managed the KAIST team. C.K prepared samples, contributed to building the experimental setup particularly the optical module, conducted measurements and analyses, verified the equations, undertook the project as PhD thesis. O.G. and C.K. discovered the initial CTRPH trapping effect behavior (in SOI) and developed the interpretation and theoretical model and. B.N. helped develop the theoretical model. O.G., C.K., B.S., B.N., D.B., J.E. and F.L. wrote the manuscript. Equations 1 to 5 were discovered and derived by O.G./B.N., B.N, O.G., O.G. and B.N/O.G. respectively. J.E. verified the theory and equations. Y.K. performed trap detection comparison. T.T. assisted in sample preparations. J.S. and M.L. developed the SOI sample. D.S.K., Y.J. and M.K. developed the perovskite samples and solar cells.

## Competing interests

This work utilized the parallel dipole line (PDL) Hall system originally developed at IBM Research and documented in the following patents: (1) O. Gunawan, T. Gokmen, US 9,041,389, (2) O. Gunawan, M. Pereira, US 9,772,385, US 9,678,040, US10,078,119 and related patents (WO 2016162772A1, UK 1717263.6, Japan 2017-552496, Germany 112016000875.9), (3) O. Gunawan, US10,197,640, (4) O. Gunawan, W. Zhou, US11,041,827.

## Additional information

Supplementary information is available for this paper.
Correspondence and requests for materials should be addressed to O.G. or B.S.

# Electronic Trap Detection with Carrier-Resolved Photo-Hall Effect

Oki Gunawan[1,†,*], Chaeyoun Kim[2,‡], Bonfilio Nainggolan[1,3], Minyeul Lee[2], Jonghwa Shin[2], Dong Suk Kim[4,5], Yimhyun Jo[4], Minjin Kim[4], Julie Euvrard[6], Douglas Bishop[1], Frank Libsch[1], Teodor Todorov[1], Yunna Kim[2], Byungha Shin[2*]

[1] *IBM T. J Watson Research Center, Yorktown Heights, NY 10598 USA*
[2] *Dept. of Materials Science and Engineering, Korea Advanced Institute of Science and Technology, Daejeon 34141, Republic of Korea*
[3] *Dept. of Physics, Arizona State University, Tempe, AZ 85287, USA*
[4] *Ulsan Advanced Energy Technology R&D Center, Korea Institute of Energy Research, Ulsan 44776, Republic of Korea*
[5] *Graduate School of Carbon Neutrality, Ulsan National Institute of Science and Technology, Ulsan 44919, Republic of Korea*
[6] *Dept. of Physics and Centre for Processable Electronics, Imperial College London, London SW7 2AZ UK*

*Corresponding authors: Oki Gunawan: ogunawa@us.ibm.com, Byungha Shin: byungha@kaist.ac.kr.

**This PDF file includes:**

Supplementary Text
Figures S1 to S7
Tables S1 to S4

**Outline:**

A. **Theory of Carrier and Trap Resolved Photo-Hall effect**
   1. Derivation
   2. The photo-Hall hyperbola equation
   3. Determination of trap density ($N_T$)
   4. Determination of trap energy level ($E_T$)
   5. Formulas for $N$-type system
B. **Carrier-Resolved Photo-Hall Formulas with Trap**
C. **Summary of Carrier and Trap Parameter Extraction using CTRPH Technique**
D. **Examples of CTRPH Analysis**
   1. Optical and physical properties of the test samples
   2. *P*-type silicon-on-insulator
   3. *N*-type silicon
   4. FAPbI$_3$ perovskite
E. **Summary of Major Trap Detection Techniques**

# A. Theory of Carrier and Trap Resolved Photo-Hall Effect

## A.1 Derivation

We will derive a general relationship of photo-Hall effect in the presence of trap that capture the role of four excitations — electric field, magnetic field, photon and phonon — in a single equation, and in the process, allowing us to extract almost all charge carrier and trap parameters in the material (see section C). The electric field (or electric current) and magnetic field excitations have been used in the traditional Hall effect which allows us to obtain the electrical conductivity $\sigma$ and Hall coefficient $H$. The key theoretical innovation in this work is the introduction of a new quantity called photo Hall conductivity (PHC) that will significantly simplify many calculations. The PHC is defined as:

$$\sigma_{PH} = \frac{\sigma^2 H}{r \mu_0}, \tag{6}$$

where $r$ is the Hall scattering factor which can be assumed to be equal to 1 in most cases [24], and $\mu_0$ is the majority carrier mobility. This concept is rooted in our earlier finding [21] that the hole-electron mobility difference can be obtained from the slope of $\sigma^2 H$ vs. $\sigma$ plot, i.e.: $\Delta\mu_H = d(\sigma^2 H)/d\sigma$, which assumes negligible trapping effect. In the presence of increasing photon intensity, $\sigma$ generally increases and $\sigma_{PH}$ will also change depending on the trapping effect and the carrier mobilities. By changing lattice temperature or phonon excitation we can alter the $\sigma_{PH}$ vs. $\sigma$ behavior that allows us to probe the trap energy level. We assume that electron and hole mobilities are constant throughout the process.

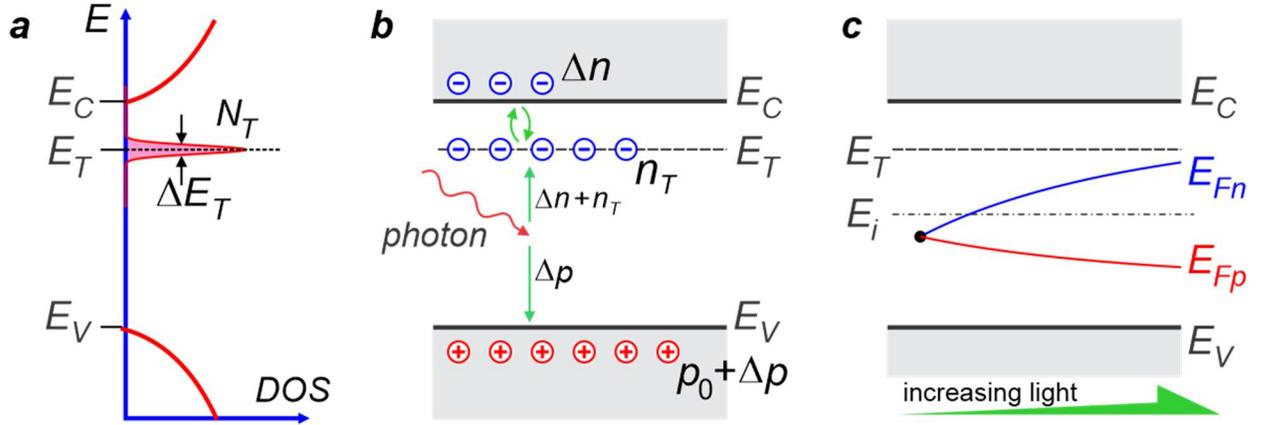

**Figure S1. A model of a single level trap in a *P*-type semiconductor with light:** **(a)** Density of states energy diagram. **(b)** The energy band diagram of a semiconductor and a single-level trap with photon excitation. **(c)** Quasi-Fermi levels for electron (blue) and hole (red) with increasing light intensity. $E_T$ and $E_i$ are the trap and intrinsic energy level respectively.

We start with single trap Shockley-Read-Hall model [30,31] for a semiconductor that describes the process of recombination and generation of charge carriers (electrons and holes) in a semiconductor via trap states [24] as shown in Fig. S1. The trap states arise from imperfections in the semiconductor crystal structure, impurities or defects. To illustrate the carrier and trap-resolved photo Hall effect calculation we a use a model based on *P*-type semiconductor, a similar model for *N*-type material is provided in section A.5. We start with a single level trap close to the conduction band (CB), thus the trap is more sensitive to electron (the minority carrier). We recognize that, in reality, there could be more than one trap present, and they could be located anywhere in the bandgap, however it is sufficient to model the most dominant trap that has the most impact to the charge-transport process, i.e., the minority carrier trap whose energy level is deep but closer to conduction band rather than the valence band (VB).

The trap has a density of state (DOS) $g_T(E)$ described by a Dirac-delta function centered at energy $E_T$ and with a total trap density $N_T$:

$$g_T(E) = N_T \delta(E - E_T). \tag{7}$$

With increasing light intensity, the Fermi level splits into electron and hole quasi Fermi levels as shown in Fig. S1c. Given the electron quasi-Fermi level $E_{Fn}$ at certain light intensity we can calculate the density of the electron in the trap $n_T$ as:

$$n_T = \int_0^\infty g_T(E) f_{FD}(E) dE = \frac{N_T}{1 + \exp((E_T - E_{Fn})/k_B T)}, \tag{8}$$

where $f_{FD}(E) = 1/[1 + \exp((E - E_F)/k_B T)]$ is the Fermi-Dirac distribution function with $E_F$ is the Fermi energy level, $k_B$ is the Boltzmann constant, and $T$ is the temperature. We further assume non-degenerate case where $E_{Fn}$ is sufficiently far away from CB ($E_C - E_{Fn} \gg k_B T$) such that Boltzmann statistic can be applied. This is a good approximation, e.g. in Si for a maximum case of $\Delta n \sim 10^{15}$/cm$^3$ under 1 sun illumination, we have $E_C - E_{Fn} \sim 10 k_B T$. We calculate $\Delta n$ using:

$$\Delta n = N_C \exp\left(-\frac{E_C - E_{Fn}}{k_B T}\right), \tag{9}$$

where $N_C$ is the effective DOS of the CB [24]:

$$N_C = 2\left(\frac{m_{d,e}^* k_B T}{2\pi \hbar^2}\right)^{3/2}, \tag{10}$$

with $m_{d,e}$* is the DOS effective mass of electron and $\hbar$ is the reduced Planck constant. By eliminating $E_{Fn}$ from Eq. (8) and (9), we obtain a nonlinear relationship between $n_T$ and $\Delta n$:

$$n_T \Delta n + N_C n_T e^{-\frac{E_{TC}}{k_B T}} - N_T \Delta n = 0, \tag{11}$$

where $E_{TC} = E_C - E_T$ is the depth of trap energy level from the CB edge (a positive number).

Next, we evaluate the conductivity and Hall coefficient in a bipolar transport process [3,32] at relatively low magnetic field: $B < 1/\mu$, where $B$ is the magnetic field and $\mu$ is the charge carrier mobility:

$$\sigma = e(\mu_N n + \mu_P p) \tag{12}$$

$$H = \frac{r(p - \beta^2 n)}{(p + \beta n)^2 e}, \tag{13}$$

where $\beta = \mu_N / \mu_P$ is the electron to hole mobility ratio, $p$ is the free hole density: $p = p_0 + \Delta p$ and $n$ is the free electron density: $n = n_0 + \Delta n \simeq \Delta n$, since $n_0 = n_i^2 / p_0 \ll \Delta n$. We also have $\Delta n$, $\Delta p$, $n_0$ and $p_0$ as the photocarrier and dark carrier density for electron and hole respectively. The charge neutrality stipulates that the photogeneration process yield equal amount of holes and electrons, with some of the electrons go to the trap: $\Delta p = \Delta n + n_T$.

Now Eq. (12) and (13) can be expressed as:

$$\sigma = \sigma_0 + e\mu_P(n_T + (1+\beta)\Delta n) \tag{14}$$

$$\sigma_{PH} = \sigma_0 + e\mu_P(n_T + (1-\beta^2)\Delta n), \tag{15}$$

where: $\sigma_0 = e\mu_0 p_0$ is the dark conductivity in the dark and $\mu_0$ is the mobility of the majority carrier in the dark. This is a linear system of two equations with two unknowns: $n_T$ and $\Delta n$, with solutions:

$$n_T = \frac{(\beta - 1)\sigma + \sigma_{PH} - \beta\sigma_0}{e\mu_P \beta} \tag{16}$$

$$\Delta n = \frac{\sigma - \sigma_{PH}}{e\mu_P \beta(\beta + 1)}. \tag{17}$$

Substituting these expressions to Eq. (11) we obtain:

$$(\beta-1)\sigma^2 + (2-\beta)\sigma\sigma_{PH} - \sigma_{PH}^2 + \beta[(\beta-1)s_E - s_N]\sigma + \beta(s_N + s_E)\sigma_{PH} - \beta^2\sigma_0 s_E = 0, \tag{18}$$

here we define the parameters:

$$s_N = e\mu_0(p_0 + N_T) \quad \text{and} \quad s_E = e\mu_0(\beta + 1)N_C \exp(-\varepsilon_{TC}), \tag{19}$$

with $\varepsilon_{TC} = (E_C - E_T)/k_B T$ is the trap energy below CB edge normalized by the thermal energy $k_B T$. As mentioned in the main text, the parameter $s_N$ and $s_E$ have very clear physical and geometrical meaning: $s_N$ depends on $N_T$ and determines the horizontal peak (or vertex) position of the hyperbola, while $s_E$ depends on $E_T$ and determines the vertical peak position (see also Fig. S3).

We note that Eq. (18) follows a generalized conic section equation [33]:

$$Q(x,y) = Ax^2 + Bxy + Cy^2 + Dx + Ey + F = 0, \tag{20}$$

where $(x, y)$ are $(\sigma, \sigma_{PH})$ and:

$$A = \beta - 1, \quad B = 2 - \beta, \quad C = -1$$
$$D = \beta[(\beta-1)s_E - s_N], \quad E = \beta(s_N + s_E), \quad F = -\beta^2 \sigma_0 s_E \tag{21}$$

We will show in the next section that Eq. (18) is guaranteed to be a hyperbola. Next, we can solve the quadratic equations in Eq. (18) and (20) to obtain $\sigma_{PH}$ as a function of $\sigma$:

$$y(x) = \left[ -Bx - E \pm \sqrt{(Bx+E)^2 - 4C(Ax^2 + Dx + F)} \right] / 2C \tag{22}$$

$$\sigma_{PH}(\sigma) = \left[ \beta(s_N + s_E) + (2-\beta)\sigma - \beta\sqrt{(s_N + s_E + \sigma)^2 - 4(s_E\sigma_0 + s_N\sigma)} \right] / 2, \tag{23}$$

which we present as Eq. 2 in the main text. The positive branch of the quadratic solution is discarded as it corresponds to negative conduction band densities $\Delta n$.

It is useful to find the two asymptotic lines of the hyperbola as ($\sigma \to \pm\infty$) from Eq. (23). They are:

$$\sigma_{PH}\big|_{\sigma \to -\infty} = \beta(s_N + s_E)/2 + \sigma \tag{24}$$

$$\sigma_{PH}\big|_{\sigma \to \infty} = \beta(s_N + s_E)/2 + (1-\beta)\sigma. \tag{25}$$

We note an important feature of this hyperbola equation, the slope $d\sigma_{PH}/d\sigma$ of the two asymptotes are 1 and 1-$\beta$ respectively. They are indicated in Fig. 1d (quadrant I and II) for the case of $P$-type materials. There is a related observation, if we calculate the slope: $s = d(\sigma^2 H)/d\sigma = r\mu_0 (d\sigma_{PH}/d\sigma)$, we will obtain $s = \mu_0$ and $s = \Delta\mu = \mu_P - \mu_N$ for the "*trap-filling*" and "*trap-saturated*" regime respectively. This means that in the first (second) regime the slope yields the majority carrier mobility (the mobility difference). This is what we observe in Fig. 1b and discussed in the main text.

We also have an interesting case in the low temperature limit where $\varepsilon_{TC}$ is large or $s_E \to 0$. Here Eq. (23) becomes a degenerate hyperbola that appears as two lines, which is given as:

$$\sigma_{PH}(\sigma) = \left[ \beta s_N + (2-\beta)\sigma - \beta\sqrt{(\sigma - s_N)^2} \right] / 2 = \left[ \beta s_N + (2-\beta)\sigma - \beta|\sigma - s_N| \right] / 2, \tag{26}$$

which can also be written as:

$$\sigma_{PH}(\sigma) = \begin{cases} \sigma & \sigma \leq s_N \\ \beta s_N + (1-\beta)\sigma & \sigma > s_N \end{cases} \tag{27}$$

The plot of this function is shown as lines DCF in Fig. S3b. This plot exhibits a peak or center point at $C(s_N, s_N)$.

## A.2 The Photo-Hall Hyperbola Equation

It is very useful to simplify the general photo-Hall equation with trap in Eq. (18), to a canonical hyperbola equation:

$$\frac{\sigma_{PH}'^2}{a^2} - \frac{\sigma'^2}{b^2} = 1. \tag{28}$$

Here we use the coordinate $(x, y)$ and its rotated coordinate $(x', y')$ to represent $(\sigma, \sigma_{PH})$ and $(\sigma', \sigma_{PH}')$ respectively. The $x'$-$y'$ axes represent the principal axes of the hyperbola. This is a hyperbola equation with vertical transverse axis (which opens toward up and down direction). The coordinate $(x', y')$ can be calculated from the translation and rotation of $(x, y)$.

We start with the basic conic section or quadratic equation [Eq. (18) and (20)] and express it in the matrix form as [33]:

$$\mathbf{x}^T A_Q \mathbf{x} = 0, \tag{29}$$

where, $A_Q$ is called the quadratic equation matrix:

$$A_Q = \begin{bmatrix} A & B/2 & D/2 \\ B/2 & C & E/2 \\ D/2 & E/2 & F \end{bmatrix} \text{ and } \mathbf{x} = \begin{bmatrix} x \\ y \\ 1 \end{bmatrix}. \tag{30}$$

Of special importance is the sub-matrix $A_{33}$, which is the characteristic matrix of the conic section. Its determinant is called the *discriminant* of the conic section [34], which will determine the type of conic section, i.e., whether it is an ellipse, parabola or hyperbola:

$$A_{33} = \begin{bmatrix} A & B/2 \\ B/2 & C \end{bmatrix}. \tag{31}$$

For our photo-Hall problem, we have:

$$|A_{33}| = AC - B^2/4 = -\beta^2/4. \tag{32}$$

Thus, it is always negative, and our conic section is guaranteed to be a hyperbola [33]. This matrix also has eigenvalues that are important parameters in this model:

$$\begin{vmatrix} A - \lambda & B/2 \\ B/2 & C - \lambda \end{vmatrix} = 0, \tag{33}$$

which yields:

$$\lambda_\pm = \left(C + A \pm \sqrt{(A-C)^2 + B^2}\right)/2 = (\beta - 2 \pm \sqrt{2\beta^2 - 4\beta + 4})/2. \quad (34)$$

There are two eigenvalues, and they only depend on $\beta$. The plot is given in Fig. S2b.

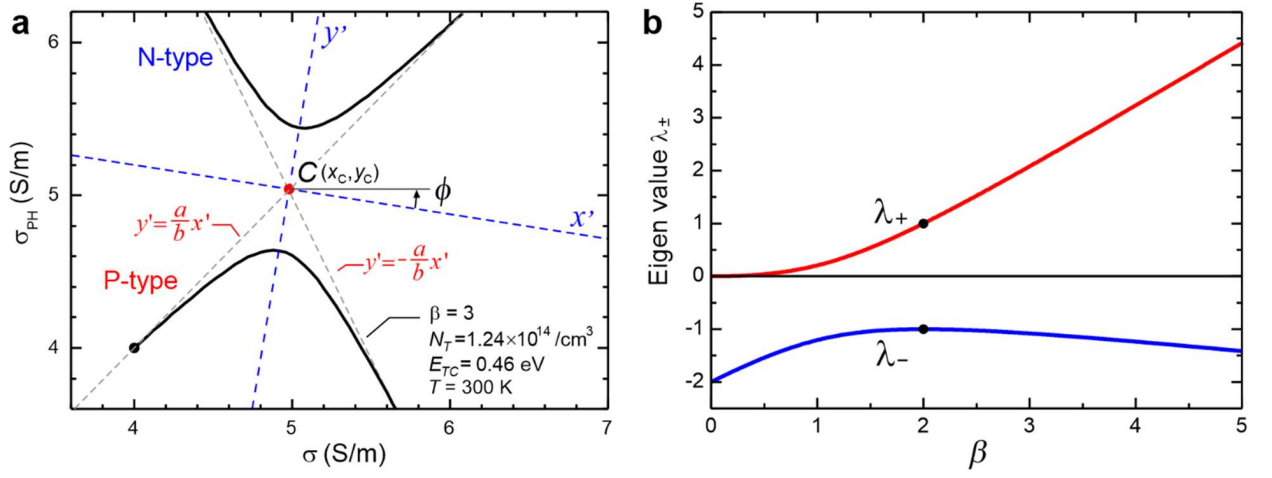

**Figure S2. (a)** The hyperbola plot and the rotated ($x'$-$y'$) coordinate frame. The lower (upper) hyperbola corresponds to a $P$-type ($N$-type) system. **(b)** Eigen values of the main hyperbola quadratic matrix as a function of $\beta$. The black dots mark the special case for $\beta = 2$ where the eigen values are: $\lambda_\pm = \pm 1$ and $\phi = 0$.

The coordinate $(x', y')$ can be calculated by applying a translation to point $C$ $(x_c, y_c)$, followed by a rotation of $-\phi$ to the original coordinate system $(x, y)$:

$$\begin{bmatrix} x' \\ y' \end{bmatrix} = \begin{bmatrix} \cos\phi & \sin\phi \\ -\sin\phi & \cos\phi \end{bmatrix} \begin{bmatrix} x - x_C \\ y - y_C \end{bmatrix}. \quad (35)$$

Point $C$ $(x_c, y_c)$ is the center point of the hyperbola which can be calculated from the condition that it is a point where the gradient of the quadratic function $Q$ in Eq. (20) vanishes [35], [36]. A way to view this problem is to consider a function $z = Q(x, y)$. If it is an ellipse, the center point would be the maximum or minimum of $z$. If it is a hyperbola, the center point would be the saddle point. In both cases, the center point is where $\nabla Q(x, y) = [0, 0]$, therefore:

$$\nabla Q(x_C, y_C) = \begin{bmatrix} \partial Q/\partial x \\ \partial Q/\partial y \end{bmatrix} = \begin{bmatrix} 0 \\ 0 \end{bmatrix} \quad \text{or} \quad \begin{bmatrix} 2A x_C + B y_C + D \\ B x_C + 2C y_C + E \end{bmatrix} = \begin{bmatrix} 0 \\ 0 \end{bmatrix}. \quad (36)$$

Now, we can solve for $x_C$ and $y_C$:

$$\begin{bmatrix} x_C \\ y_C \end{bmatrix} = \frac{1}{4AC - B^2} \begin{bmatrix} BE - 2CD \\ BD - 2AE \end{bmatrix} = \begin{bmatrix} s_N - s_E \\ s_N + (\beta - 1)s_E \end{bmatrix}. \tag{37}$$

This solution which defines the coordinate of center point C, plays a very important role in this model as it allows us to determine trap density $N_T$ and trap energy level $E_T$.

The semi major and semi minor axis $a$ and $b$ are given as [33]:

$$a^2 = -\frac{|A_Q|}{\lambda_+ \lambda_-^2} \quad \text{and} \quad b^2 = -\frac{|A_Q|}{\lambda_+^2 \lambda_-}, \tag{38}$$

where $|A_Q|$ is the determinant of $A_Q$: $|A_Q| = (4ACF - AE^2 - B^2F + BDE - CD^2)/4$. We define:

$$K = \frac{|A_Q|}{\lambda_+ \lambda_-} = \frac{CD^2 - BDE + AE^2}{B^2 - 4AC} + F = \beta^2 s_E(s_N - \sigma_0) = e\mu_0 \beta^2 N_T s_E. \tag{39}$$

Note that the value $K$ is always positive. Now we have:

$$a = \sqrt{-K/\lambda_-} \quad \text{and} \quad b = \sqrt{K/\lambda_+}. \tag{40}$$

Next, we calculate the rotation angle $\phi$ of the hyperbola's principal axes as shown in Fig. S2a. We can derive this angle in a simple way by noting that the principal axes of the hyperbola bisects the asymptotes. We can determine the asymptotes of the hyperbola as $x, y \to \pm\infty$, thus only the first three terms dominate in the hyperbola equation in Eq. (20):

$$Ax^2 + Bxy + Cy^2 = 0. \tag{41}$$

We can obtain the asymptote line equations by solving for $y$:

$$y = \frac{-B \pm \sqrt{B^2 - 4AC}}{2C} x, \tag{42}$$

which implies that the slope of the asymptotes are: $m_{1,2} = (-B \pm \sqrt{B^2 - 4AC})/2C$. From Fig. S2a, we see that the principal axes form an angle $\phi$ with respect to the $x$-axis which bisects the two asymptotes. In other words, $\phi$ is the average of the angles of the two asymptotes: $\phi_1$ and $\phi_2$. we have:

$$2\phi = \phi_1 + \phi_2, \tag{43}$$

where $\tan \phi_{1,2} = m_{1,2}$. Therefore, we have:

$$\tan 2\phi = \tan(\phi_1 + \phi_2) = \frac{\tan\phi_1 \tan\phi_2}{1-\tan\phi_1\tan\phi_2} = \frac{m_1+m_2}{1-m_1 m_2} = \frac{B}{A-C} = \frac{2-\beta}{\beta}, \text{ or:} \quad (44)$$

$$\phi = \tfrac{1}{2}\tan^{-1}(2/\beta - 1) \quad (45)$$

We also note another important characteristic, the ratio of *b* and *a* determines the curvature or the eccentricity (ε) of the hyperbola. The eccentricity of the hyperbola is given as [34]:

$$\varepsilon = \sqrt{1+(b/a)^2}, \quad (46)$$

which depends on ratio of *b* and *a*, and in our CRPH problem, it solely depends on β:

$$\left(\frac{b}{a}\right)^2 = \sqrt{\frac{-\lambda_-}{\lambda_+}} = \frac{\beta - 2 - \sqrt{2\beta^2 - 4\beta + 4}}{\beta - 2 + \sqrt{2\beta^2 - 4\beta + 4}}. \quad (47)$$

The eccentricity, ranging from 1 to ∞, determines the curvature of the hyperbola. The hyperbola is sharp when ε is low (~1) or β is large, and the hyperbola is broad when ε is large or β is very small ($\to 0$).

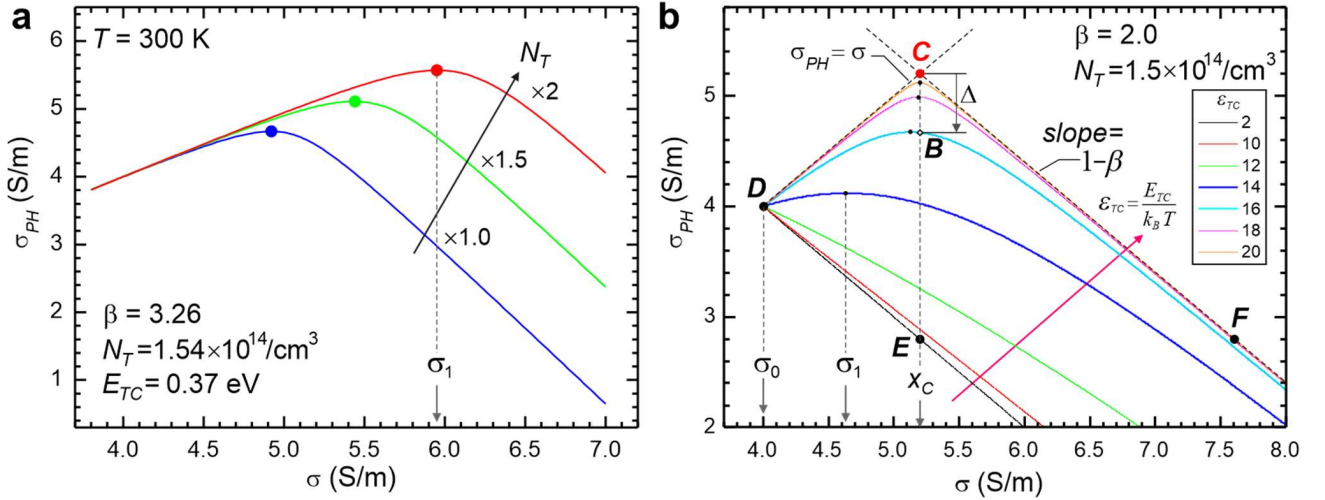

**Figure S3. Simulation of the impact of trap parameter $N_T$ and $E_T$ in a *P*-type material.** (a) Effect of increasing $N_T$ that shifts the hyperbola peak to the right at fixed temperature. (b) Effect of increasing $\varepsilon_T$ that makes the hyperbola sharper.

In summary, we have expressed the relationship between $\sigma_{PH}$ vs. $\sigma$ as a compact hyperbola equation in Eq. (28), and elaborate its parameters such as $x_C$, $y_C$, $\phi$, *a* and *b* in terms of the physical parameters β, $N_T$ and $E_T$. In Fig. S3 we simulate a set of hyperbola curves where $N_T$ and $E_T$ vary. We make few notable remarks about the photo-Hall hyperbola characteristics:

1. Increasing $N_T$ will shift the hyperbola peak (or vertex) to the right as shown in Fig. S3a, as it will take higher light intensity and photo carrier density to fill all the traps, after which the curve starts to turn around.

2. Increasing $\varepsilon_{TC}$, which occur when $E_{TC}$ is larger or $T$ is lower, will make the hyperbola sharper (see Fig. S3b) and approaching the two asymptotic lines. This is what we observe in Fig. 2 for the *p*-SOI data, where the $\sigma_{PH} - \sigma$ hyperbola curves get sharper at lower temperature.

3. The center point $C$ plays a crucial role to determine $N_T$ and $E_T$ (see section A.3 and A.4). Point $C$ is determined from the intersection of $\sigma_{PH} = \sigma$ line and the second asymptote line. This implies that we need to take enough data points at high light intensity to obtain the second asymptote. We note that the line: $\sigma_{PH} = \sigma$ is not necessarily the first asymptote at high temperature, but it becomes the first asymptote at low temperature.

4. If there is no trap, or the trapping effect is negligible, i.e., $N_T = 0$ or $E_T = 0$, the first segment of the hyperbola curve is absent, instead we only have the second segment with slope $1 - \beta$ (e.g. line DE in Fig. S3b).

5. We note a special case when $\beta = 2$. From Eq. (45) we have $\phi = 0$, in other words, the principal axes of the hyperbola is already aligned with the *x* and *y* axes and thus no rotation is needed and $(\sigma', \sigma'_{PH})$ is equal to $(\sigma, \sigma_{PH})$. This is also consistent with the fact that the slopes of the two asymptotes are: 1 and $1 - \beta = -1$ respectively, implying that the hyperbola is already symmetric in the original axes. We also have the eigen values: $\lambda_{\pm} = \pm 1$ and thus $a = b = K$. In this case, the $\sigma_{PH} - \sigma$ relationship is reduced to a very simple equation: $\sigma_{PH}^2 = \sigma^2 + K^2$.

6. The hyperbola bending is the sharpest when the minority carrier mobility is larger than that of the majority (i.e., quadrant I and IV in Fig. 1d). This is reasonable, because once the trap is full and the minority carrier start to contribute to the transport, the change in $\sigma_{PH}$ is more significant if the minority carrier mobility is larger.

7. It is fascinating that the rich physics of the photo-Hall effect with trap can be captured by a succinct, hyperbola equation. Conic section solutions such as ellipse and hyperbola are well known in physics to describe motions of celestial objects. Recently, the hyperbolic trajectory of asteroid "1I/'Oumuamua", characterized by its unprecedentedly high eccentricity ($\varepsilon = 1.2$), played a decisive role in confirming its groundbreaking identification as the first known interstellar object to visit our solar system, as reported by Meech *et al.* in Nature [37].

## A.3 Determination of Trap Density ($N_T$)

From the hyperbola model, we can determine the trap density in a simple way using the peak (or the vertex) coordinate. The x-coordinate of the peak is close to that of center point $C$: $x_C = s_N - s_E$. We note that, in the limit: $\varepsilon \to \infty$, which occurs at low temperature, the hyperbola becomes very sharp and we have: $s_E \sim 0$ and $x_C$ becomes close to the peak position: $x_C \simeq \sigma_1$. Using Eq. (19) we have $\sigma_1 = s_N = \sigma_0 + e\mu_0 N_T$. Therefore we can solve for $N_T$ and obtain the simple formula below. We denote $\tilde{N}_T$ as the estimated trap density.

$$\tilde{N}_T = \frac{\sigma_1 - \sigma_0}{e\mu_0}. \tag{48}$$

**Temperature-dependent effect in $N_T$ determination:**

Strictly speaking, the $N_T$ determination from the hyperbola peak (or inflection point) using Eq. (48) contains some temperature-dependent factor, which can be derived below. The peak position can be calculated from Eq. (23):

$$\left.\frac{d\sigma_{PH}}{d\sigma}\right|_{\sigma=\sigma_1} = 1 - \frac{\beta}{2} - \frac{\beta(\sigma - s_N + s_E)}{2\sqrt{\sigma^2 - 2(s_N - s_E)\sigma + (s_N + s_E)^2 - 4\sigma_0 s_E}} = 0 \tag{49}$$

$$\sigma_1^2 - 2(s_N - s_E)\sigma_1 + (s_N + s_E)^2 - 4\sigma_0 s_E + \frac{\beta^2}{1-\beta}e\mu_0 N_T s_E = 0. \tag{50}$$

Solving this quadratic equation, we obtain, the peak position:

$$\sigma_1 = (s_N - s_E) - (\beta - 2)\sqrt{\frac{e\mu_0 N_T s_E}{\beta - 1}}. \tag{51}$$

So now we have the exact expression for the $\tilde{N}_T$:

$$\tilde{N}_T = \frac{\sigma_1 - \sigma_0}{e\mu_0} = N_T - (1+\beta)N_C \exp(-\varepsilon_{TC}) - |\beta - 2|\sqrt{\frac{\beta+1}{\beta-1}N_T N_C \exp(-\varepsilon_{TC})}. \tag{52}$$

We note that the effective density of states $N_C$ contains temperature dependent factor $T^{3/2}$. We can write this equation as:

$$\tilde{N}_T(T) = N_T - a_1 u(T) - a_2 \sqrt{u(T)}, \tag{53}$$

where $u(T) = T^{3/2} \exp(-\varepsilon_{TC})$ is the temperature dependent factor and constants: $a_1 = 2(\beta+1)(m_{d,e}^* k_B / 2\pi\hbar^2)^{3/2}$ and $a_2 = |\beta - 2|\sqrt{(\beta+1)N_T/(\beta-1)} \times (m_{d,e}^* k_B / 2\pi\hbar^2)^{3/4}$. This equation predicts that $\tilde{N}_T = N_T$ at low temperature and as the temperature increases $\tilde{N}_T$ will drop. This effect is observed in Fig. 2b where $\tilde{N}_T$ drops at higher temperature, and the actual $N_T$ can be determined from the lowest temperature data.

## A.4 Determination of Trap Energy Level ($E_T$)

**Trap energy level determination from a single temperature measurement:**

We can determine the trap energy level from the gap parameter "$\Delta$". We note that $\Delta$ is very close to the semimajor axis $a$, in fact, if $\phi = 0$, the gap is exactly the same as the hyperbola semi-major axis, i.e., $\Delta = a$. We can calculate the maximum gap $\Delta_{max}$ from the center point $C$ is $(s_N, s_N)$. The maximum gap $\Delta_{max}$, (segment CE) can be calculated in Fig. S3b as:

$$\Delta = y_C - \sigma_{PH}(x_C) = \beta\sqrt{s_E(s_N - \sigma_0)} = \beta\sqrt{e\mu_0 N_T s_E}. \tag{54}$$

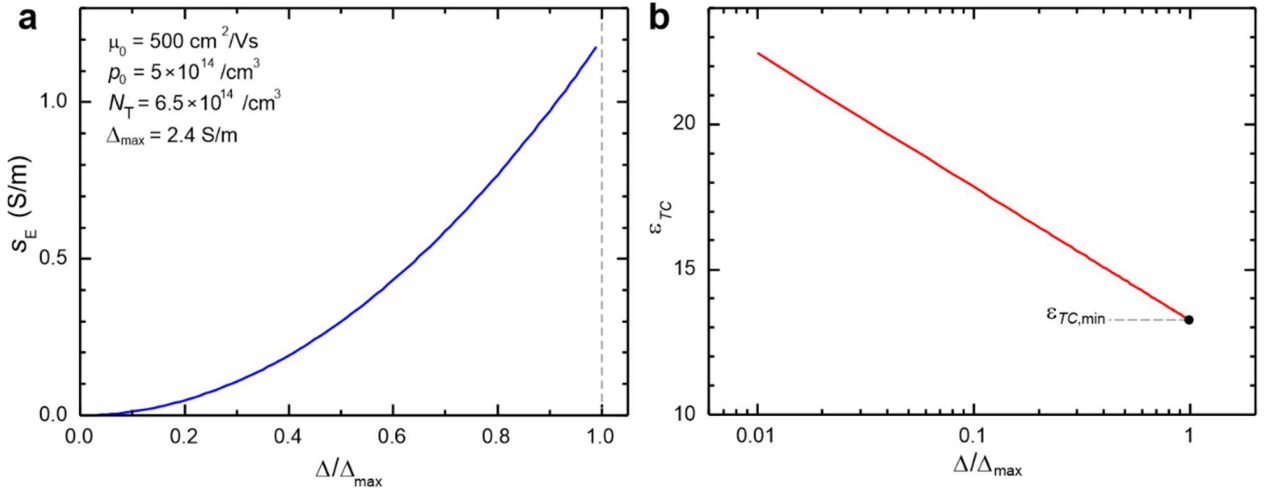

**Figure S4. Determination of $E_T$ from the gap $\Delta$.** (a) $s_E$ vs. $\Delta$ plot. (b) Determination of $\varepsilon_{TC}$ from $\Delta$.

The maximum gap $\Delta_{max}$ can be obtained from the segment CE in Fig. S3b. Point $C$ is obtained when $\varepsilon \to \infty$ or $s_E = 0$. Point E is from segment DE which occurs when $\varepsilon \to 0$ or $s_E \gg s_N$. In this limit Eq. (23) reduces to:

$$\tilde{\sigma}_{PH}(\sigma) = (1-\beta)\sigma + \beta\sigma_0, \tag{55}$$

which is the equation for line DE. We obtain a simple expression:

$$\Delta_{max} = y_C - y_E = s_N - \tilde{\sigma}_{PH}(s_N) = e\beta\mu_0 N_T. \tag{56}$$

We can calculate the parameter $s_E$ as:

$$s_E = \frac{\Delta^2}{e\beta^2 \mu_0 N_T}. \tag{57}$$

From Eq. (19), we can solve for $E_T$ as a function of $\Delta$:

$$E_{TC} = k_B T \ln\left(\frac{e\mu_0(1+\beta)N_C}{s_E}\right) = k_B T \ln\left(\frac{e^2\mu_0^2 \beta^2 (1+\beta) N_C N_T}{\Delta^2}\right). \tag{58}$$

If we substitute $\Delta = \Delta_{\max}$ we obtain the minimum trap energy level that can be detected as shown in Fig. S4b which is given as:

$$E_{TC,\min} = k_B T \ln\left[(1+\beta)N_C / N_T\right]. \tag{59}$$

This equation implies that: (1) there is a minimum energy level that we can detect at a given temperature, (2) we can detect smaller trap energy level at lower temperature. We can plot $s_E$ vs. $\Delta/\Delta_{\max}$ from Eq. (57) as shown in Fig. S4a, we see that $s_E$ increases monotonically with $\Delta$. The plot of trap energy vs. $\Delta$ is shown in Fig. S4b and the $\varepsilon_{TC,\min}$ is also indicated.

**Trap energy level determination from variable temperature measurements:**

Besides using the gap "$\Delta$", we can also utilize temperature-dependent slope of $S = d\sigma_{PH}/d\sigma$ in the dark to extract $E_T$. We start from a series expansion of $\sigma_{PH}(\sigma)$ in terms of $\sigma - \sigma_0$:

$$\sigma_{PH}(\sigma) = \sigma_0 + \left[1 - \frac{\beta s_E}{s_N + s_E - \sigma_0}\right](\sigma - \sigma_0) + O\left[(\sigma - \sigma_0)^2\right]. \tag{60}$$

We define the slope in the dark $S_0 = d\sigma_{PH}/d\sigma\big|_{\sigma=\sigma_0}$ as:

$$S_0 = \frac{d\sigma_{PH}}{d\sigma}\bigg|_{\sigma=\sigma_0} = 1 - \frac{\beta(\beta+1)N_C \exp(-\varepsilon_{TC})}{N_T + (\beta+1)N_C \exp(-\varepsilon_{TC})}. \tag{61}$$

We can rearrange this expression and isolate the exponential factors, and express the effective DOS [Eq. (10)] in terms of its temperature-dependent factor, i.e., $N_C = N_0 T^{1.5}$:

$$\frac{N_0 T^{1.5}}{N_T}(\beta+1)\left[\frac{\beta}{1-S_0} - 1\right] = \exp(\varepsilon_{TC}) \tag{62}$$

$$\frac{1-S_0}{(\beta+1)(\beta-1+S_0)T^{1.5}} = \frac{N_0}{N_T}\exp(-\varepsilon_{TC}) \tag{63}$$

$$\ln\left(\frac{1-S_0}{(\beta+S_0-1)(1+\beta)T^{1.5}}\right) = c_0 - \frac{E_{TC}}{k_B T}, \tag{64}$$

where $c_0 = \ln(N_0/N_T)$. Therefore, Eq. (64) allows us to extract the trap energy level via temperature-dependent measurement using Arrhenius-type analysis. To use this equation, we need a set of temperature-dependent measurements of slope $S_0$ and $\beta$ from the second asymptote. An example of this analysis is presented in Fig. 2 and its corresponding discussion in the main text.

## A.5 Formulas for *N*-type System

The behavior of *N*-type semiconductors mirrors that of the *P*-type equations presented in section A.1. In general we can obtain the equations for the *N*-type materials from those of the *P*-type by using the following transformation: $H \to -H$, $\sigma_{PH} \to -\sigma_{PH}$, $\beta \to 1/\beta$, and $N_C \to N_V$. Here the minority carrier is hole, and the trap energy level is close to the valence band (VB). The trapped hole density is given as:

$$p_T = \int_0^\infty g_T(E)[1 - f_{FD}(E)]dE = \frac{N_T}{1 + \exp[(E_{Fp} - E_T)/k_B T]}, \tag{65}$$

and the hole photocarrier density in the VB is:

$$\Delta p = N_V \exp[-(E_{Fp} - E_V)/k_B T], \tag{66}$$

where $N_V$ is the DOS of VB [24]:

$$N_V = 2\left(\frac{m_{d,p}^* k_B T}{2\pi \hbar^2}\right)^{3/2}, \tag{67}$$

with $m_{d,p}^*$ is the hole DOS effective mass. Similar to Eq. (11), the non-linear equation relating the trapped hole and free hole is given as:

$$p_T \Delta p + N_V p_T e^{-\frac{E_{TV}}{k_B T}} - N_T \Delta p = 0, \tag{68}$$

where $E_{TV} = E_T - E_V$ is the trap energy level from the valence band edge $E_V$ (a positive number). The transport equations for the *N*-type materials can be written as:

$$\sigma = \sigma_0 + e\mu_N(p_T + (1 + 1/\beta)\Delta p) \tag{69}$$

$$\sigma_{PH} = -\sigma_0 - e\mu_N(p_T + (1 - 1/\beta^2)\Delta p). \tag{70}$$

From these equations we can solve for $p_T$ and $\Delta p$:

$$p_T = \frac{(1-\beta)\sigma - \beta\sigma_{PH} - \sigma_0}{e\mu_N} \tag{71}$$

$$\Delta p = \frac{\beta^2(\sigma + \sigma_{PH})}{e\mu_N(1+\beta)}. \tag{72}$$

The explicit hyperbola equation in Eq. (19) for *N*-type material becomes:

$$\sigma_{PH}(\sigma) = -\left[s_N + s_E + (2\beta - 1)\sigma + \sqrt{(s_N + s_E + \sigma)^2 - 4(s_E\sigma_0 + s_N\sigma)}\right]/2\beta, \quad (73)$$

where: $s_N = e\mu_0(n_0 + N_T)$ and $s_E = e\mu_0(1 + 1/\beta)N_V e^{-\varepsilon_{TV}}$.

For $N_T$ determination, the equation for *N*-type material is the same as shown in Eq. (48), but for $E_T$ determination, it is given as:

$$E_{TV} = k_B T \ln\left(\frac{e\mu_0(1+1/\beta)N_V}{s_E}\right) = k_B T \ln\left(\frac{e^2\mu_0^2(1+\beta)N_V N_T}{\beta^3 \Delta^2}\right). \quad (74)$$

For the temperature-dependent version like in Eq. (64), we have:

$$\ln\left(\frac{1+S_0}{(1-S_0-\beta)(1+\beta)T^{1.5}}\right) = c_0 - \frac{E_{TV}}{k_B T} \quad (75)$$

## B. Carrier-Resolved Photo-Hall Formulas with Trap

Once we solve for the trapping effect as described by the hyperbola model in Section A, we occasionally need to solve for full carrier-resolved solutions if the mobilities vary, which often occurs at high light intensity. For clarity, we revisit the CRPH solutions in the presence of trap and also address the variable mobility situation when the traps are assumed to be full ($n_T \sim N_T$). They are largely similar to our earlier CRPH model [21], except we include the role of $n_T$ in the solutions.

Here, we attempt to solve for three unknowns: $\Delta p, \mu_P$ and $\mu_N$ from three experimentally measured quantities: $\sigma$, $H$, and $\Delta\mu = d(\sigma^2 H)/d\sigma$. We consider the hole and electron density: $p = p_0 + \Delta p$, $n = n_0 + \Delta n \simeq \Delta n$, and the charge neutrality condition: $\Delta p = \Delta n + n_T$. Starting from the bipolar Hall transport equations in Eqs. (12) and (13), we have:

$$\sigma = e[(p_0 + \Delta p)\mu_P + (\Delta p - n_T)(\mu_P - \Delta\mu)] \quad (76)$$

$$\sigma^2 H = e[\mu_P^2(p_0 + \Delta p) - (\mu_P - \Delta\mu)^2(\Delta p - n_T)]. \quad (77)$$

Solving for $\mu_P$ and $\Delta p$, we have:

$$\mu_P = \frac{1}{2}\left[\Delta\mu - \sqrt{\frac{4\sigma(\sigma H - \Delta\mu) + e\Delta\mu^2(n_T + p_0)}{e(p_0 + n_T)}}\right] \quad (78)$$

$$\Delta p = \frac{\sigma + e n_T (\mu_P - \Delta\mu) - p_0 \mu_P e}{e(2\mu_P - \Delta\mu)}. \tag{79}$$

Then we can obtain the full solutions: $\mu_N = \mu_P - \Delta\mu$ and $\Delta n = \Delta p - n_T$. For the calculation when the trap is full (e.g. at high light intensity), we can use: $n_T \sim N_T$ and thus $\Delta n = \Delta p - N_T$.

As explained in the previous work ([21], SI section B.2), if the mobilities vary – which could occur at high injection ($\Delta p, \Delta n \gg p_0$), the actual mobility difference will be modified. We rederive again these results and simplify them:

$$\Delta\mu = \left[\frac{d(\sigma^2 H)}{d\sigma} + c_1\right] c_2, \quad \text{where:} \tag{80}$$

$$c_1 = 2e\left(n\mu_N \frac{d\mu_N}{d\sigma} - p\mu_P \frac{d\mu_P}{d\sigma}\right) \tag{81}$$

$$c_2 = \left(1 - e\left[n\frac{d\mu_N}{d\sigma} + p\frac{d\mu_P}{d\sigma}\right]\right)^{-1}. \tag{82}$$

We note that if the mobilities are constant, Eq. (80) reduces to the basic form: $\Delta\mu = d(\sigma^2 H)/d\sigma$ as expected. If the initial solution of the CTRPH exhibits significant variation in the mobilities then we need to employ this "variable-mobility correction" in iterative fashion, until the solution converges [21]. The correction tends to increase $\Delta\mu$ at high light intensity, as we found in the perovskite calculation in Fig. S7a.

## C. Summary of Carrier and Trap Parameter Extraction using CTRPH Technique

Here we summarize all the parameters that can be calculated from the CTRPH technique in two lists: charge carrier parameters and trap parameters. These include nearly all properties that we want to know from charge carriers and trap in semiconductor. The entries in yellow mark the primary parameters, that come as direct results from the experimental measurements. The rest are secondary parameters which mean they can be derived from the primary parameters. In general, the key data in the CTRPH experiment is a set of $\sigma$, $H$ and $G_\gamma$ measurements at various light intensities, and they can be repeated as a function of temperature $T$. Our theoretical model, as described in section A, could yield all the parameters below for both $P$ and $N$-type semiconductors (with some slight differences in the formulas between the two).

**C.1 Charge carrier parameters:**

| No | Parameter | Description | Formula for P or N-type system | |
|---|---|---|---|---|
| 1. | $\Delta n$ | electron photocarrier density | P | $\Delta n = \dfrac{\sigma - \sigma_{PH}}{e\mu_P \beta(\beta+1)}$ |
| | | | N | $\Delta n = \Delta p + p_T$ |
| 2. | $\Delta p$ | hole photocarrier density | P | $\Delta p = \Delta n + n_T$ |
| | | | N | $\Delta p = \dfrac{\beta^2(\sigma + \sigma_{PH})}{e\mu_N(1+\beta)}$ |
| 3. | $n_T$ or $p_T$ | trapped electron or hole density | P | $n_T = \dfrac{(\beta-1)\sigma + \sigma_{PH} - \beta\sigma_0}{e\mu_P \beta}$ |
| | | | N | $p_T = \dfrac{(1-\beta)\sigma - \beta\sigma_{PH} - \sigma_0}{e\mu_N}$ |
| 4. | $\mu_N$ | electron mobility | P | $\mu_N = \beta \mu_P$ |
| | | | N | $\mu_N = \mu_0$ (dark) |
| 5. | $\mu_P$ | hole mobility | P | $\mu_P = \mu_0$ (dark) |
| | | | N | $\mu_P = \mu_N / \beta$ |
| 6. | $\tau_N$ | electron recombination lifetime | $\tau_N = \Delta n / G$ | |
| 7. | $\tau_P$ | hole recombination lifetime | $\tau_P = \Delta p / G$ | |
| 8. | $D_N$ | electron diffusion coefficient | $D_N = k_B T \mu_N / e$ | |
| 9. | $D_P$ | hole diffusion coefficient | $D_P = k_B T \mu_P / e$ | |
| 10. | $D_A$ | ambipolar diffusion coefficient | $D_A = \dfrac{n+p}{n/D_P + p/D_N}$ | |
| 11. | $L_{D,N}$ | electron diffusion length | $L_{D,N} = \sqrt{D_N \tau_N}$ | |
| 12. | $L_{D,P}$ | hole diffusion length | $L_{D,P} = \sqrt{D_P \tau_P}$ | |
| 13. | $L_{D,A}$ | ambipolar diffusion length | $L_{D,A} = \sqrt{D_A \tau_A}$ | |
| 14. | $QF_N$ | electron quasi-Fermi level | $QF_N = k_B T \ln(n/n_i)$ | |
| 15. | $QF_P$ | hole quasi-Fermi level | $QF_P = k_B T \ln(p/n_i)$ | |
| 16. | $QFLS$ | quasi-Fermi level splitting | $QFLS = QF_N + QF_P$ | |
| 17. | $\eta$ | ideality factor | $\eta = (dQFLS / d\ln G_\gamma)/k_B T$ | |

**Table S1.** The charge carrier parameters extracted using CTRPH method. Parameters in yellow indicate primary extraction results.

**Notes:**

(1) In the CTRPH model (entry #1-5), we assume constant mobilities (and thus $\beta$) and constant dark carrier density $p_0$ or $n_0$, which implies there should not be any persistent photoconductivity effect (see e.g. [38]). At high light intensity the mobilities could vary, and one could use variable mobility model as described in previous work [21] and section B.

(2) For lifetime calculation (#6 and 7) we assume 100% quantum efficiency thus generation rate $G$ and the absorbed photon density $G_\gamma$ are equal: $G = G_\gamma$. We also assume 100% exciton dissociation efficiency.

(3) For ambipolar diffusion length (#13), we can calculate the effective lifetime as: $\tau_A = (n+p)/(n/\tau_N + p/\tau_P)$, using Matthiessen's rule of carrier lifetimes weighted by their respective densities.

(4) Given electron and hole densities, the quasi Fermi levels and the quasi Fermi level splitting (QFLS) can also be calculated as suggested in Ref. [20] and shown in entry #14, 15 and 16. $n_i$ is the intrisinc carrier density [24]. The quasi Fermi level is calculated with respect to the intrinsic Fermi level $E_i$. The QFLS is useful to estimate the potential open circuit voltage that a material could deliver in a solar cell application [39].

(5) The ideality factor $\eta$ can also be calculated [20], which is obtained from general relationship that relates QFLS with photon flux $\Phi$ [40]: $\Phi = \Phi_0 \exp(QFLS/\eta k_B T)$. Since $\Phi$ is proportional to the absorbed photon density $G_\gamma$, we can derive $\eta$ as shown in entry #17.

(6) We can also analyze the carrier density as a power law of light intensity, i.e.,: $\Delta n = G_\gamma^m$. Evaluating the exponent factor $m$ is useful to analyze the recombination mechanisms (see e.g. [41,42]). At high light intensiity limit, where $\Delta n \simeq \Delta p$, the exponent factor $m$ is related to the ideality factor $\eta$ as: $m = \eta/2$. This can be derived as follows:

$$QFLS = k_B T \ln(\Delta n^2 / n_i^2), \qquad dQFLS/d\ln\Delta n = 2k_B T \qquad (83)$$

$$\eta = \frac{dQFLS}{d\ln G_\gamma} \frac{1}{k_B T} = \frac{dQFLS}{d\ln \Delta n} \frac{d\ln \Delta n}{d\ln G_\gamma} \frac{1}{k_B T} = 2m. \qquad (84)$$

For example, many high performance solar cells has $\eta \sim 1$ [43], which implies: $m \sim 0.5$ (at high light intensity) as we observe in our perovskite study (Fig. 3 and [21]).

## C.2 Trap parameters:

| No | Parameters | Description | Formula |
|---|---|---|---|
| 1. | $N_T$ | trap density | $\tilde{N}_T = \dfrac{\sigma_1 - \sigma_0}{e\mu_0}$ |
| 2. | $E_T$ | trap energy level | $E_{TC} = k_B T \ln\left(\dfrac{e^2 \mu_0^2 \beta^2 (\beta+1) N_C N_T}{\Delta^2}\right)$ <br><br> P <br><br> $\ln\left(\dfrac{1-S_0}{(\beta+S_0-1)(1+\beta)T^{1.5}}\right) = c_0 - \dfrac{E_{TC}}{k_B T}$ |

| | | | | |
|---|---|---|---|---|
| | | | N | $E_{TV} = k_B T \ln\left(\dfrac{e^2 \mu_0^2 (1+\beta) N_V N_T}{\beta^3 \Delta^2}\right)$ <br><br> $\ln\left(\dfrac{1+S_0}{(1-S_0-\beta)(1+\beta)T^{1.5}}\right) = c_0 - \dfrac{E_{TV}}{k_B T}$ |
| 3. | $\sigma_N$ | electron scattering cross section | P | $\sigma_N = 1/v_{Tn} N_T \tau_{N,SRH}$ |
| | | | N | $\sigma_N = 1/v_{Tn} n_T \tau_{N,SRH}$ |
| 4. | $\sigma_P$ | hole scattering cross section | P | $\sigma_P = 1/v_{Tp} n_T \tau_{P,SRH}$ |
| | | | N | $\sigma_P = 1/v_{Tp} N_T \tau_{P,SRH}$ |

**Table S2.** The trap parameters extracted using CTRPH method. Parameters in yellow indicate primary extraction results.

**Note:**

(1) For best $N_T$ and $E_T$ extraction, trap detection can be obtained when the $\sigma_{PH} - \sigma$ bending is significant, i.e., the minority mobility is larger than that of majority and sufficiently low temperature measurement ($E_{TC} \gg k_B T$).

(2) For scattering cross section calculation (#3 and 4), we use Schottky-Read-Hall (SRH) recombination model [44]. Considering a *P*-type material as an example, we have the rate equations where SRH process dominate:

$$dn/dt = G - n\sigma_N v_{Tn}(N_T - n_T) \tag{85}$$

$$dp/dt = G - p\sigma_P v_{Tp} n_T. \tag{86}$$

At very low light intensity where the trap is nearly empty ($n_T \ll N_T$) we have:

$$\tau_{N,SRH} = \frac{\Delta n}{G} = \frac{1}{\sigma_N v_{Tn} N_T} \quad \text{or} \quad \sigma_N = \frac{1}{v_{Tn} N_T \tau_{N,SRH}} \tag{87}$$

$$\tau_{P,SRH} = \frac{\Delta p}{G} = \frac{1}{\sigma_P v_{Tp} n_T} \quad \text{or} \quad \sigma_P = \frac{1}{v_{Tp} n_T \tau_{P,SRH}}, \tag{88}$$

where $v_T = \sqrt{8k_B T/\pi m^*}$ is the thermal velocity of electron or hole and $m^*$ is their corresponding effective mass [45]. Note that, using example of *P*-type system, for the $\sigma_P$ calculation, since $n_T$ may be inaccurate at low light intensity, we can use the rate calculation: $1/\tau_P$ vs. $n_T$, i.e., $\sigma_P = [d(1/\tau_P)/dn_T]/v_{Tp}$.

# D. Examples of Carrier and Trap Resolved Photo-Hall Effect Analysis

## D.1 Optical and physical properties of the test samples

| No | Parameter | Symbol | Unit | Sample | | |
|---|---|---|---|---|---|---|
| | | | | p-SOI | n-Si | FAPbI$_3$ perov- |
| | **Sample:** | | | | | |
| 1. | Thickness | d | μm | 5.0 | 280 | 0.7 |
| 2. | Active area (width x length) | W x L | mm | 1 x 2 | 2 x 4 | 2 x 4 |
| 3. | Reflectivity | R | | 0.334 | 0.352 | 0.171 |
| 4. | Absorption coefficient | α | /cm | 3.72x10$^3$ | 3.66x10$^3$ | 4.43x10$^3$ |
| 5. | DOS effective mass | $m_e^*/m_h^*$ | | 1.08/0.55 | 1.08/0.55 | 0.23/0.29 |
| 6. | Intrinsic carrier density | $n_i$ | /cm$^3$ | 1.1x10$^{10}$ | 1.1x10$^{10}$ | 1.57x10$^5$ |
| | **PDL Hall System:** | | | | | |
| 7. | Light wavelength | λ | μm | 615 | 615 | 615 |
| 8. | Reference PD QE | $QE_{REF}$ | | 0.765 | 0.765 | 0.765 |
| 9. | PD calibration factor | $k_{PD}$ | | 0.70 | 0.62 | 0.57 |
| 10. | Optical setup constant | $k_G$ | /Acm$^3$s | 8.58x10$^{22}$ | 1.56x10$^{21}$ | 6.77x10$^{23}$ |

**Table S3.** The device and optical parameters for the samples used in this work. PD is photodetector and QE is quantum efficiency. Perovskite DOS effective mass is obtained from Ref. [46].

We present the physical, optical and system parameters for all three samples that we used in this work as shown in Table S3 above.

**Notes:**

(1) Sample active area is the area defined by the four Hall bar terminals that are used in the longitudinal and transverse resistance measurement.

(2) The effective mass is used for effective DOS calculation ($N_C$ and $N_V$), intrinsic carrier density ($n_i$) for quasi-Fermi levels calculation and thermal velocity ($v_T$) for the trap scattering cross section.

(3) In the experiment, we monitor the light intensity by measuring the current in the monitor photodetector cell ($I_{PD-MON}$) (see Fig. 1a). The calculation detail is also presented in our previous work ([21], SI section C). The optical properties of the films are needed to calculate the absorbed photon density given as: $G_\gamma = k_G(\lambda) I_{PD,MON}$, where $k_G$ is the optical setup constant that is wavelength-dependent:

$$k_G(\lambda) = \frac{k_{PD}[1-R(\lambda)]}{e\, QE_{REF}(\lambda)\, A_{REF}} \frac{[1-\exp(-\alpha(\lambda)d)]}{d}, \tag{89}$$

where $k_{PD}$ is the light calibration factor as measured by photodetector (PD), i.e., the ratio between the reference and monitor PD: $k_{PD} = I_{PD,REF}/I_{PD,MON}$, averaged thorughout all operating light intensity.

(4) The light intensity (in W/m²) impinging on the sample is:

$$I_L = \frac{I_{PD,MON} k_{PD}}{QE_{REF} A_{REF}} \frac{hc}{e\lambda}. \tag{90}$$

where $A_{REF}$ =7.5 mm² is the area of the reference PD, $h$ is the Planck constant, and $c$ is the speed of light. This formula is used to calculate the maximum light intensity impinging on the sample (e.g. see Fig. 1 and 3).

## D.2 Carrier and Trap Resolved Photo-Hall Analysis of *p*-SOI sample

Here we present the complete CTRPH analysis of the *p*-SOI sample that is discussed in the main text. The physical and optical parameters of this sample are shown in Table S3. First, we obtain the photo-Hall data and perform the curve fitting as discussed in the main text and presented again in Fig. S5a. From the dark Hall data point, we have: $\sigma$ = 3.98 S/m, $p_0$ = 5.65×10¹⁴ /cm³ and $\mu_P$ = 440 cm²/Vs. The curve-fitting using the hyperbolic model in Eq. (23) yields: $\beta$ = 3.26, $N_T$ = 1.54×10¹⁴ /cm³ and $E_T$ = 0.46 eV. Here we obtain the electron mobility: $\mu_N = \beta \mu_P$=1432 cm²/Vs which is close to the known values of mobilities in silicon (1500 cm²/Vs [3]).

We use Eq. (8), (9) and (23) to calculate $\Delta n$, $n_T$ and thus obtain $n = \Delta n$ and $p = \Delta n + n_T$ vs. $G_\gamma$ as shown in Fig. S5b. Unlike in our earlier work, where we neglect the trapping effect, now we can completely resolve the electron and hole properties and thus calculate the mobility, lifetime and diffusion lengths for both of them as shown in Fig. S5b-g. We also report the quasi-Fermi levels and its ideality factor. In addition, using the formulas in Table S2, we can calculate the scattering cross section factors which are: $\sigma_N = 1.81 \times 10^{-2}$ nm² and $\sigma_P = 3.72 \times 10^{-3}$ nm² for electron and hole respectively.

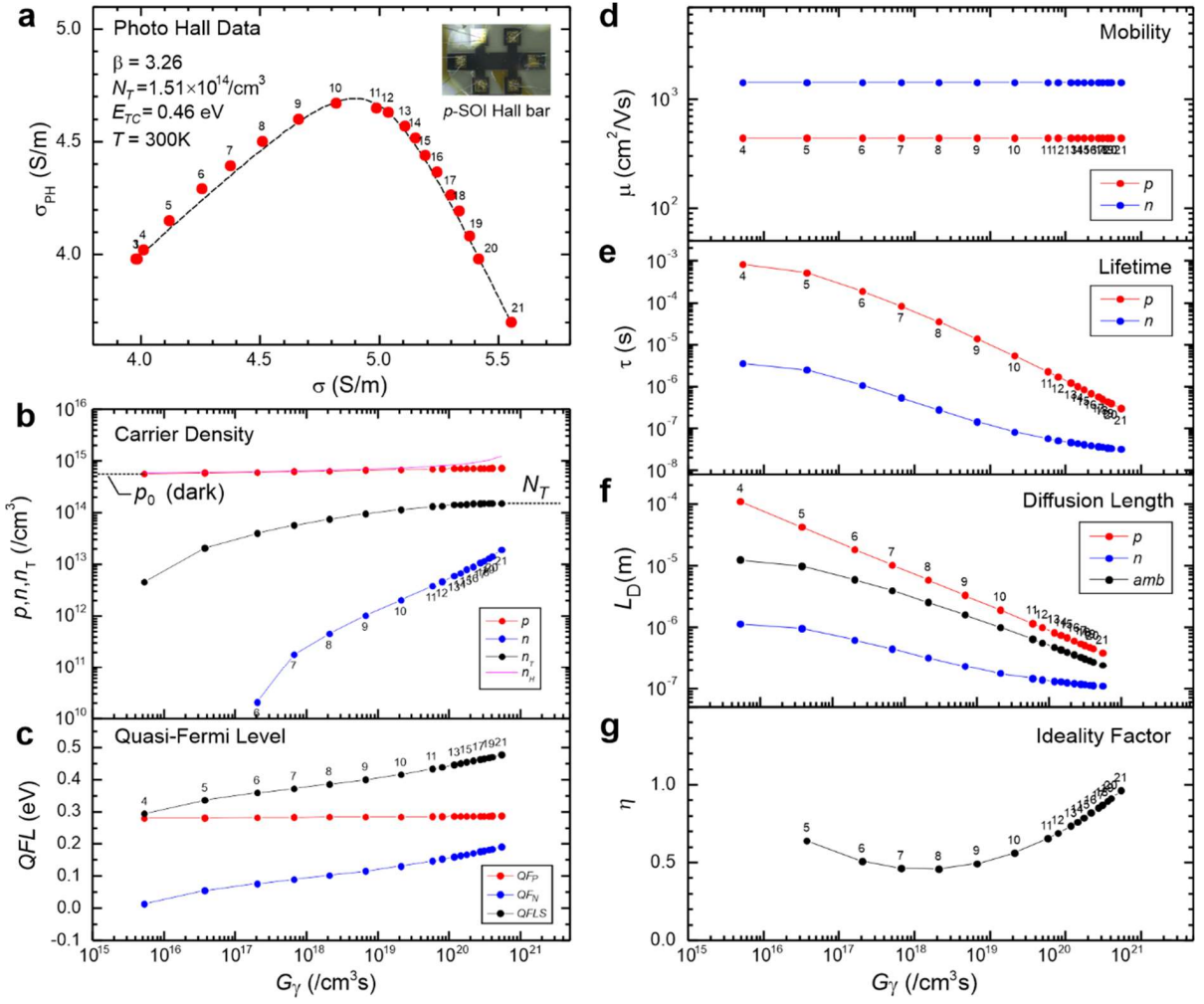

**Figure S5. Complete CRPH analysis of the *p*-SOI sample: (a)** The photo Hall data with laser light λ= 615 nm and maximum intensity $I_L$ =157 mW/cm$^2$. **Inset:** the *p*-SOI sample. **(b)** Carrier density for *p*, *n* and $n_T$ vs. absorbed photo density $G_\gamma$, and $n_H = 1/eH$ is the Hall density. **(c)** Quasi Fermi level. **(d)** Mobility. **(e)** Recombination lifetime. **(f)** Diffusion length. **(g)** Ideality factor.

### D.3 Carrier and Trap Resolved Photo-Hall Analysis of *n*-Si sample

We present the CTRPH study in an *N*-type single crystal silicon as shown in Fig. S6. The physical and optical properties of the sample is shown in Table S3. From the dark Hall data point, we have: $\sigma_0$ = 0.0982 S/m, dark electron density: $n_0$ = 3.58×10$^{12}$ /cm$^3$, and mobility $\mu_N$ = 1703 cm$^2$/Vs. We plot the photo Hall data: $\sigma_{PH}$ vs. $\sigma$ as shown in Fig. S6a. We observe the plot exhibits a bending behavior expected for an *N*-type system with $\mu_P < \mu_N$, i.e., the quadrant III behavior in Fig. 1d, where there are two line segments with negative slopes and a weak bending. We determine the inflection point at $\sigma_1$ = 0.167 S/m, and using Eq. 3, we obtain trap density $N_T$ = 2.49×10$^{12}$ /cm$^3$.

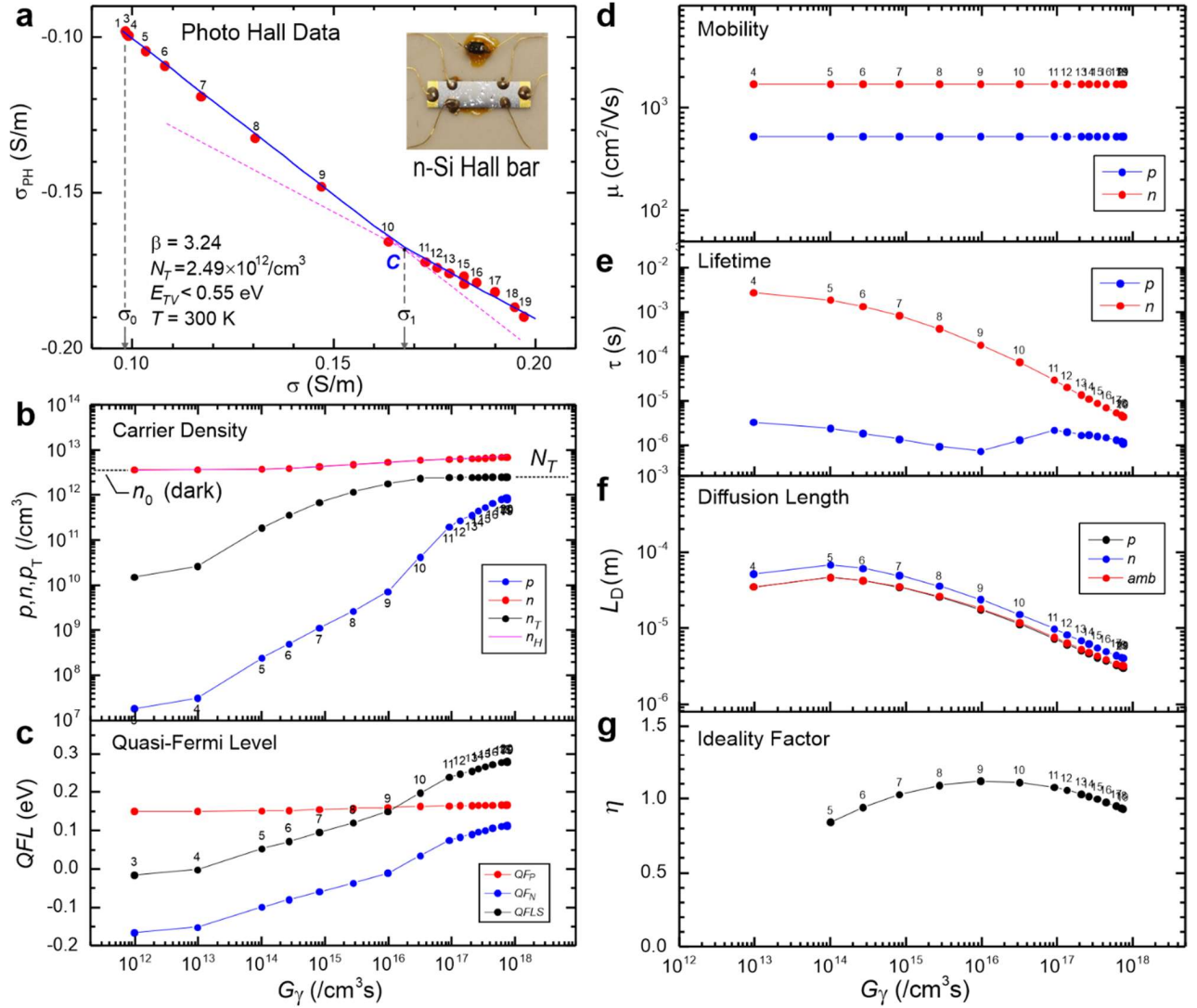

**Figure S6. Complete CRPH analysis of an *N*-type Si sample: (a)** The photo Hall data with laser light λ= 615 nm and maximum intensity $I_L$ =10.5 mW/cm$^2$. Blue curve: hyperbola fit from Eq. (23). **Inset:** the *N*-type Si Hall sample. **(b)** Carrier density vs. absorbed photo density ($G_\gamma$). **(c)** Quasi-Fermi levels. **(c)** Mobility. **(d)** Recombination lifetime. **(e)** Diffusion length. **(g)** Ideality factor.

From the second segment, we obtain the slope: $S = -1 + 1/\beta = 0.691$, which yield $\beta = 3.24$ and thus hole (minority) mobility of: $\mu_P = \mu_N / \beta = 526$ cm$^2$/Vs. We note that, again, both mobilities are consistent with the known values in silicon of $\mu_P \sim 500$ and $\mu_N \sim 1500$ cm$^2$/Vs [3] and similar to the results from the *p*-SOI sample in this work. For the trap energy level, due to very weak bending near the inflection point and insufficient data points, we can not determine the parameter Δ accurately. However we can establish its upper limit, i.e., $\Delta < 9.4 \times 10^{-4}$ and thus upper bound of the trap energy, i.e., $E_{TV} < 0.55$ eV.

The full set of CRPH analysis outcome including carrier density, quasi-Fermi levels, mobility, lifetime, diffusion lengths and ideality factors are presented in Fig. S6b-g. Furthermore, from the lifetime data near dark, we obtain scattering cross section: $\sigma_N = 1.1 \times 10^{-2}$ nm$^2$ and $\sigma_P = 1.49 \times 10^{-3}$ nm$^2$ for electron and hole respectively.

## D.4 Carrier and Trap Resolved Photo-Hall Analysis of the FAPbI$_3$ Perovskite Sample

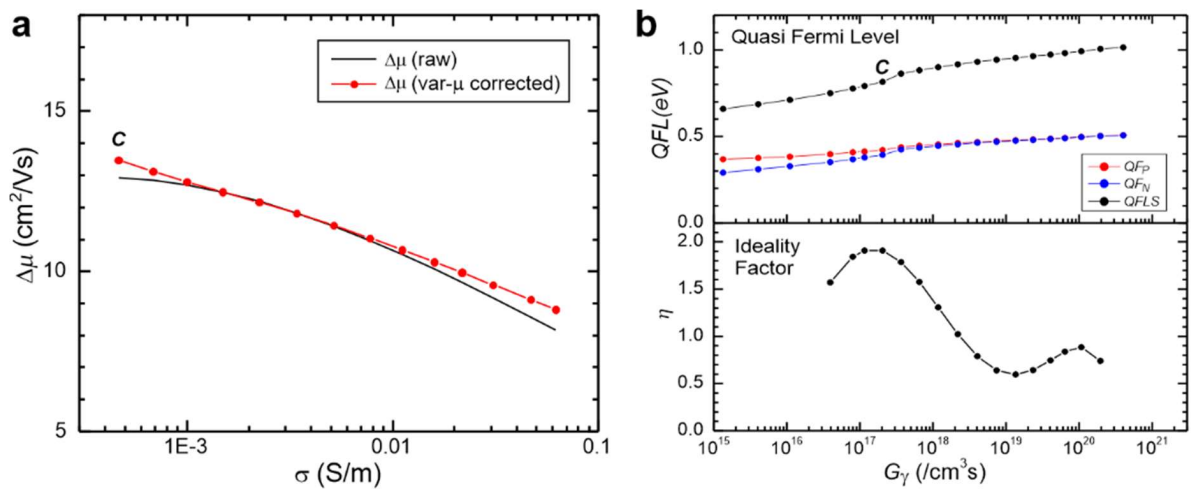

**Figure S7. Additional analysis on the FAPbI$_3$ perovskite sample: (a)** The mobility difference ($\Delta\mu$) before and after variable-mobility correction. **(b)** Quasi Fermi level and ideality factor as a function of absorbed photon density $G_\gamma$.

We present the details of the FAPbI$_3$ perovskite CTRPH analysis. The initial objective is to obtain a set of primary quantities: $\mu_P, \mu_N, \Delta n, n_T$ and $\Delta p$. As mentioned in the text, we divide the analysis to two regimes: (1) Trapping regime (segment *D-C-E*) and (2) Trap-full or high light intensity regime (segment *E-F*). In the first regime, the $\sigma_{PH} - \sigma$ plot shows a weak bending, consistent with the quadrant II behavior in Fig. 1d, where the majority mobility is larger than that of minority ($\mu_P > \mu_N$). Fortunately, we could observe an inflection point at *C*, which allows us to extract the trap density using Eq. (3), i.e., $N_T = 1.9 \times 10^{12}$ /cm$^3$. The segment *C-E* of the curve, where the trap is expected to be full, yields a slope *S*=0.81 and thus $\beta \sim 0.19$. The mobility in the dark is $\mu_P \sim (12 \pm 2)$ cm$^2$/Vs. As the bending is weak, we can not reliably determine the gap "Δ", however using simulation of the hyperbola equation in Eq. (23), we estimate $E_T \sim 0.3\text{-}0.7$ eV. Then we can calculate $\Delta n$, $n_T$ and $\Delta p$ using equations in Table S1 and plotted in Fig. 3b.

Next, we analyze the second regime at high light intensity. To solve the photo-Hall transport problem, we first calculate the mobility difference using: $\Delta\mu = (2 + d\ln H/d\ln\sigma)\sigma H$, which is more appropriate for high injection regime like in this perovskite sample [21]. To facilitate a smooth derivative calculation (for $\Delta\mu$ and variable-mobility correction), we fit $H$ vs. $\sigma$ data with a smooth second-degree polynomial in the log-log scale. To ensure continuous solution, we start the analysis from point C, where the trap is approximately full. We obtain the (raw) $\Delta\mu$ data which is not constant as shown in Fig. S7a. Then we apply the variable mobility correction using Eq. (80) in iterative fashion until we obtain self-consistent $\Delta\mu$ solution as shown in Fig. S7b (red curve). Finally, we solve for: $\Delta n$, $n_T$ and $\Delta p$ using Eqs. (78) and (79) and other charge carrier parameters as shown in Fig. 3. In addition, we also present the Quasi-Fermi Level plots and its ideality factor in Fig. S7b. We note that, at maximum light intensity (~16 mW/cm$^2$), the QFLS is ~1.0 V which is close to $V_{OC}$ ~ 1.07 V of the corresponding solar cell at the same intensity ([27], Fig. 2). Finally, from the lifetime data near dark, we obtain scattering cross section: $\sigma_N = 0.39\,\text{nm}^2$ and $\sigma_P = 0.75\,\text{nm}^2$ for electron and hole respectively. We also make another comment, in the high light intensity regime $E$-$F$ (Fig. 3b), we observe: $\Delta n \simeq \Delta p$, thus one could also solve the problem while ignoring the trapping effect like in our previous report [21].

# E. Summary of Major Trap Detection Techniques

| No. | Technique | Trap properties | | | Carrier* properties | Notes |
|---|---|---|---|---|---|---|
| | | $N_T$ | $E_T$ | $\sigma_{P,N}$ | | |
| 1. | Deep-Level Transient Spectroscopy (DLTS) [10,47] | √ | √ | √ | X | Need junction. Could detect multiple trap levels, energy profile, has high sensitivity. |
| 2. | Drive-Level Capacitance Profiling (DLCP) [1,48] | √ | √ | X | X | Need junction. Could yield trap spatial information. |
| 3. | Space Charge Limited Current (SCLC) [14,49,50] | X | √ | X | $\mu$ | Can detect hole or electron trap density. Drift majority mobility can be extracted. |
| 4. | Transient photo-luminescence (TrPL) [16] | √ | X | √ | $\tau$ | Requires modeling to extract trap density $N_T$ and $\Delta n$. |
| 5. | Thermal Admittance Spectroscopy (TAS) [2,15] | √ | √ | √ | X | Need junction. Yield energy profile. |
| 6. | Thermally Stimulated Current (TSC) [51,52] | √ | √ | X | X | May fail detect traps due to incomplete trap filling and partial detrapping during thermalization. $N_T$ can only be estimated. |
| 7. | Transient Photo-conductivity (TRPC) or Microwave Conductivity (TRMC) [17,53] | √ | X | √ | $\mu_p+\mu_n$, $\tau$ | Modeling required to extract $N_T$. |

| 8. | Transient Photo Hall spectroscopy [19] | √ | √ | √ | τ | Need to have high mobility to yield strong transient signal. More carrier properties can be extracted but not attempted. |
|---|---|---|---|---|---|---|
| 9. | Photo Hall effect spectroscopy [18] | X | √ | √ | X | Sub-bandgap light needed. $\sigma_{N/P}$ measured by complementary transient photo-conductivity. More carrier properties can be extracted but not attempted. |
| 10. | Constant Light-Induced Magneto Transport/Photo Hall (CLIMAT) [20] | √ | ~ | √ | $\Delta n$, $\Delta p$, $n_T$, $\mu_N$, $\mu_P$, $\tau_N$, $\tau_P$ | $N_T$ is determined from simulation and fitting. $E_T$ is estimated. Yield nearly all carrier-resolved charge carrier parameters. |
| 11. | Carrier and Trap Resolved Photo Hall effect (CTRPH/this work) | √ | √ | √ | $\Delta n$, $\Delta p$, $n_T$, $\mu_N$, $\mu_P$, $\tau_N$, $\tau_P$ | $N_T$ and $E_T$ are determined from exact formulas, including $T$-dependent equation for $E_T$. Yield nearly all carrier-resolved charge carrier parameters. |

**Table S4.** Major trap detection techniques and their output capability. $\sigma_{P/N}$ is the capture cross section for hole or electron. Marker √, ~, and × indicate capable to measure, approximate and not capable respectively. *Only primary carrier properties outputs are listed. There could be more secondary carrier parameters that can be calculated (e.g. $D$, $L_D$ and quasi-Fermi levels), see section C for a complete list.

Table S4 summarizes various major trap detection techniques and their output capabilities, including some comments. The cited references are mostly application examples in perovskite materials. The CTRPH technique yields the most comprehensive list of charge and carrier trap parameters output as much as 17×$N$ charge carrier and 4 trap parameters in total, with $N$ is the number of light intensity settings. Best results are obtained when if the $\sigma_{PH} - \sigma$ plot has strong hyperbolic bending (low eccentricity).